\documentclass[letterpaper, 10 pt, conference]{ieeeconf}  

\IEEEoverridecommandlockouts                              

\usepackage{graphics} 
\usepackage{amsmath} 
\usepackage{amssymb}  
\usepackage{romannum} 
\usepackage{color} 
\usepackage{paralist}
\usepackage{graphicx}  
\usepackage{subfigure}
\usepackage{multirow} 
\usepackage{booktabs} 
\usepackage{siunitx} 
\usepackage{changes}

\newtheorem{exmp}{Example}
\newtheorem{theorem}{Theorem}

\newtheorem{rem}{Remark}
\newtheorem{proposition}{Proposition}
\newtheorem{assumption}{Assumption}

\newcommand{\boldxi}{\boldsymbol{\xi}}
\newcommand{\ud}{\operatorname{d}\!}

\title{\LARGE \bf
A Polynomial Chaos Approach to Robust $\mathcal{H}_\infty$ Static Output-Feedback Control with Bounded Truncation Error}

\author{Yiming Wan, Dongying E. Shen, Sergio Lucia, Rolf Findeisen, and Richard D. Braatz
\thanks{This work is supported by the National Natural Science Foundation of China under Grant No. 61803163, and the DARPA Make-It program under contract ARO W911NF-16-2-0023.}
\thanks{Yiming Wan is with School of Artificial Intelligence and Automation, Huazhong University of Science and Technology, and Key Laboratory of Image Processing and Intelligent Control, Ministry of Education, Wuhan 430074, China. Email address:{\tt\small ywan@hust.edu.cn}}
\thanks{Dongying E. Shen and Richard D. Braatz are with Massachusetts Institute of Technology, 77 Massachusetts Avenue, Cambridge, MA 02139, USA. Email address: 
	{\tt\small \{dongying, braatz\}@mit.edu}}
\thanks{Sergio Lucia is with TU Dortmund University, 44227 Dortmund, Germany. Email address: {\tt\small sergio.lucia@tu-dortmund.de}}
\thanks{Rolf Findeisen is with Otto-von-Guericke University Magdeburg, 39106 Magdeburg, Germany.
	Email address:{\tt\small rolf.findeisen@ovgu.de}}
}

\begin{document}

\maketitle
\thispagestyle{empty}
\pagestyle{empty}

\begin{abstract}
This article considers the $\mathcal{H}_\infty$ static output-feedback control for linear time-invariant uncertain systems with polynomial dependence on probabilistic time-invariant parametric uncertainties. 
By applying polynomial chaos theory, the control synthesis problem is solved using a high-dimensional expanded system which characterizes stochastic state uncertainty propagation. 
A closed-loop polynomial chaos transformation is proposed to derive the closed-loop expanded system.
The approach explicitly accounts for the closed-loop dynamics and preserves the $\mathcal{L}_2$-induced gain, which results in smaller transformation errors compared to existing polynomial chaos transformations. 
The effect of using finite-degree polynomial chaos expansions is first captured by a norm-bounded linear differential inclusion, and then addressed by formulating a robust polynomial chaos based control synthesis problem.
This proposed approach avoids the use of high-degree polynomial chaos expansions to alleviate the destabilizing effect of truncation errors, which significantly reduces computational complexity. 
In addition, some analysis is given for the condition under which the robustly stabilized expanded system implies the robust stability of the original system. 
A numerical example illustrates the effectiveness of the proposed approach.
\end{abstract}

\section{Introduction}
Numerous studies have been reported on robust control design subject to model uncertainties, e.g., see \cite{petersen2014robust} and references therein. 
The widely used worst-case strategy tends to produce highly conservative performance because the worst-case scenario may have vanishingly low probability of occurrence. In contrast to a worst-case performance bound, practical interest in the performance variation or dispersion across the uncertainty region has motivated recent research on probabilistic robustness \cite{petersen2014robust}. The design objective either adopts a probability-guaranteed worst-case performance bound \cite{Tempo2013,yaesh2003probability}, or optimizes the averaged performance, at the expense of an increased worst-case performance bound \cite{boyarski2005robust}. This line of research considers polytopic uncertainties \cite{boyarski2005robust,yaesh2003probability}, or affine dependence on multiplicative white noises \cite{hinrichsen1998stochastic}. The randomized algorithm proposed in \cite{Tempo2013} can address nonlinear uncertainty structures, but can be computationally demanding since a large number of samples is often required.

As mentioned above, the probabilistic description of parametric uncertainties allows optimizing the averaged performance.
A common assumption is multiplicative white noise as in \cite{hinrichsen1998stochastic}, with arbitrarily fast and unbounded rate of parameter changes, which is unrealistic for many applications. 
Often, uncertain parameters change significantly slower than the underlying system dynamics. 
For this reason, they can be regarded as time invariant, and their
probabilistic information can be obtained from a priori knowledge or parameter identification.
When accounting for the time invariance of uncertain parameters, 
uncertainty propagation of the system state becomes non-Markovian  \cite{Paulson2015}, thus its computation usually resorts to computationally expensive sampling-based approaches \cite{Tempo2013}.
Recently, some progress in robust control synthesis has been made by exploiting polynomial chaos (PC) theory as a non-sampling approach for uncertainty propagation, e.g., see \cite{Fisher2009,Hsu2020design,nandi2017poly,paulson2018efficient,Shen2017}.
The PC based stochastic Galerkin method characterizes the evolution of probability distributions of the underlying stochastic system states by a high-dimensional transformed deterministic system that describes the evolution dynamics of the polynomial chaos expansion (PCE) coefficients. 
Then the control synthesis problem can be solved by using the PCE-transformed system.
An important consideration in PCE-based control is that due to using finite-degree PCEs, the PCE-transformed system has an inevitable approximation error, which may destabilize the closed-loop dynamics and reduce closed-loop performance
\cite{audouze2016priori,Lucia2017,muhlpfordt2018comments}.
Although increasing the PCE degree can reduce the effect of PCE truncation errors, computational complexity increases significantly, as the state dimension of the PCE-transformed system factorially grows with the PCE degree.

In this paper, a PCE-based $\mathcal{H}_\infty$ static output-feedback (SOF) control approach is proposed for uncertain linear time-invariant (LTI) systems with polynomial dependence on probabilistic time-invariant parametric uncertainties.
The proposed PCE-based transformation explicitly copes with the closed-loop dynamics, and preserves the $\mathcal{L}_2$-induced gain, which results in a smaller approximation error compared to existing PCE-transformed systems reported in \cite{Shen2017,Wan2018ACC}.
To account for the perturbations from the PCE truncation errors, 
a norm-bounded linear differential inclusion (LDI) is constructed for the PCE-transformed system. 
Then a robust PCE-based synthesis problem is formulated to minimize an upper bound of the $\mathcal{L}_2$-induced gain from disturbances to the controlled outputs. This formulation enforces closed-loop stability and robustifies the control performance against PCE truncation errors. 
These benefits are achieved without the high computational complexity resulted from using a high-degree PCE.

This paper is organized as follows. Section \ref{sect:prob} states the robust $\mathcal{H}_\infty$ SOF control problem. Section \ref{sect:pce} reviews preliminaries on PC theory. Section \ref{sect:pce_cldyn} derives the PCE-transformed closed-loop system. In Section \ref{sect:sof_synthesis}, a robust PCE-based SOF control synthesis method is proposed by using the norm-bounded LDI to account for the PCE truncation errors. A simulation study and some concluding remarks are given in Sections \ref{sect:sim} and \ref{sect:conclusion}, respectively.

Notations: For a continuous-time vector-valued signal $x(t)$, $\left\| x(t) \right\|_{\mathcal{L}_2}$ denotes the $\mathcal{L}_2$ norm defined as $\left( \int_0^\infty x^\top(t) x(t) \text{d} t \right)^{{1}/{2}}$. 
The 2-norm of a vector $x$ is defined as $\left\| x \right\|_2 = \sqrt{ x^\top x}$.
For a square matrix $X$, let $\text{He}\{X\}$ represent $X + X^\top$. 
For a symmetric matrix $X$, its positive and negative definiteness is denoted by $X>0$ and $X<0$, respectively.
The vectorization of a matrix $X$ is denoted by $\text{vec}(X)$.
The Kronecker product is represented by $\otimes$.

\section{Problem Statement}\label{sect:prob}
Consider an uncertain LTI system described by 
\begin{subequations}\label{eqn:open_sys}
	\begin{align}
	\dot{\mathbf{x}}(t,\boldsymbol{\xi}) &=                          \mathbf{A}(\boldsymbol{\xi})\mathbf{x}(t,\boldsymbol{\xi})  + \mathbf{B}_{\mathbf{w}}(\boldsymbol{\xi})\mathbf{w}(t)  + \mathbf{B}(\boldsymbol{\xi})\mathbf{u}(t,\boldsymbol{\xi}) \label{eqn:dynamics}\\
	\mathbf{z}(t,\boldsymbol{\xi})  &= \mathbf{C}_{\mathbf{z}}\mathbf{x}(t,\boldsymbol{\xi}) + \mathbf{D}_\mathbf{zw}  \mathbf{w}(t) + \mathbf{D}_\mathbf{z}\mathbf{u}(t,\boldsymbol{\xi})  \label{eqn:outputz}\\
	\mathbf{y}(t,\boldsymbol{\xi})  &= \mathbf{C}(\boldsymbol{\xi})\mathbf{x}(t,\boldsymbol{\xi})  + \mathbf{D}_{\mathbf{w}}(\boldsymbol{\xi})\mathbf{w}(t) \label{eqn:outputy}
	\end{align}
\end{subequations}
where $\mathbf{x} \in \mathbb{R}^{n_x}$ is the state, $\mathbf{u} \in \mathbb{R}^{n_u}$ is the control input, $\mathbf{w} \in \mathbb{R}^{n_w}$ is the external disturbance, $\mathbf{y} \in \mathbb{R}^{n_y}$ is the measured output, $\mathbf{z} \in \mathbb{R}^{n_z}$ is the controlled output, and $\boldxi \in \mathbb{R}^{n_\xi}$ is the uncertain parameter vector that lies within a bounded set $\Xi$. 
In the closed-loop system, the system state $\mathbf{x}$, control input $\mathbf{u}$, measured output $\mathbf{y}$, and controlled output $\mathbf{z}$ all depend on $\boldsymbol{\xi}$. The matrices $\mathbf{C}_\mathbf{z}$, $\mathbf{D}_\mathbf{zw}$, and $\mathbf{D}_\mathbf{z}$ in (\ref{eqn:outputz}) are independent of  $\boldsymbol{\xi}$, because they are determined by the performance specifications in the controlled output $\mathbf{z}$. 

\begin{assumption}\label{ass:polymat}
The system matrices $\mathbf{A}$, $\mathbf{B}_\mathbf{w}$, $\mathbf{B}$, $\mathbf{C}$, and $\mathbf{D}_\mathbf{w}$ in (\ref{eqn:open_sys}) are  polynomial functions of $\boldxi$.
\end{assumption}

The above assumption of polynomial dependence on $\boldxi$ is not restrictive since non-polynomial nonlinear dependence can be accurately approximated by polynomials or piecewise polynomials \cite{Suli2003}.

The objective of this paper is to design an SOF controller 
	\begin{equation}\label{eqn:sofcontroller}
	\mathbf{u}(t,\boldsymbol{\xi}) = \mathbf{K} \mathbf{y} (t,\boldsymbol{\xi})
	\end{equation}
that solves the $\mathcal{H}_{\infty}$-optimal control problem
\begin{equation}\label{eq:prob}
\begin{aligned}
\min_{\mathbf{K},\gamma>0}& \quad \gamma  \\
\text{s.t.}&\quad \mathbb{E}_{\boldxi}\!\left\{ \left\| \mathbf{z}(t,\boldxi) \right\|_{\mathcal{L}_2}^2 \right\}
\leq \gamma^2 \left\| \mathbf{w}(t) \right\|_{\mathcal{L}_2}^2, \\
& \quad \forall \left\| \mathbf{w}(t) \right\|_{\mathcal{L}_2} < \infty
\end{aligned}
\end{equation}
for the closed-loop system
\begin{subequations}\label{eq:clsys}
\begin{align}
\mathbf{\dot x}(t,\boldxi) &= \mathbf{A}_\text{cl}(\boldxi)\mathbf{x}(t,\boldxi)+
\mathbf{B}_{\text{cl}}(\boldxi)\mathbf{w}(t) \label{eq:clsys_dyn}\\
\mathbf{z}(t,\boldxi)&= \mathbf{C}_{\text{cl}}(\boldxi)\mathbf{x}(t,\boldxi) + \mathbf{D}_\text{cl}(\boldxi)\mathbf{w}(t) \label{eq:clsys_out}
\end{align}
\end{subequations}
with 
\begin{equation}\label{eq:clsysmatdef}
\begin{aligned}
\mathbf{A}_{\text{cl}}(\boldxi) &= \mathbf{A}(\boldxi)+\mathbf{B}(\boldxi)\mathbf{K}\mathbf{C}(\boldxi), \\
\mathbf{B}_{\text{cl}}(\boldxi) &= \mathbf{B}_\mathbf{w}(\boldxi) + \mathbf{B}(\boldxi)\mathbf{K}\mathbf{D}_\mathbf{w}(\boldxi), \\ 
\mathbf{C}_\text{cl}(\boldxi) &= \mathbf{C}_\mathbf{z} + \mathbf{D}_\mathbf{z}\mathbf{K}\mathbf{C}(\boldxi), \\
\mathbf{D}_\text{cl}(\boldxi) &= \mathbf{D}_\mathbf{zw} + \mathbf{D}_\mathbf{z}\mathbf{K}\mathbf{D}_\mathbf{w}(\boldxi).
\end{aligned}
\end{equation}
The $\mathcal{L}_2$-induced gain $\gamma$ from the disturbance $\mathbf{w}(t)$ to the controlled output $\mathbf{z}(\boldxi,t)$ generalizes the $\mathcal{H}_{\infty}$ norm of a deterministic LTI system.
The mathematical expectation $\mathbb{E}_{\boldxi}\!\left\{ \cdot \right\}$ in \eqref{eq:prob} accounts for the time-invariant probabilistic parametric uncertainties $\boldsymbol{\xi}$.

To explicitly account for the polynomial dependence on probabilistic uncertain parameters $\boldxi$, the basic idea of solving the problem \eqref{eq:prob} is: 
(\romannum{1}) first apply the PCE-based transformation to derive a 
high-dimensional transformed closed-loop system that describes the dynamics of PCE coefficients; and (\romannum{2}) then construct a $\mathcal{H}_\infty$-control synthesis problem for the PCE-transformed closed-loop system.
However, due to using finite-degree PCEs, 
PCE truncation errors lead to discrepancy between the PCE-transformed closed-loop system and the original closed-loop system.  
Even when the PCE-transformed closed-loop system is stabilized,
the presence of PCE truncation errors can destabilize the original closed-loop system or reduce its closed-loop performance \cite{Lucia2017}. Most PCE-based control literature have not addressed this challenge, which motivates this investigation in how to
(\romannum{1}) perform the PCE-based transformation to achieve small transformation errors;
(\romannum{2}) take into account the PCE truncation errors in solving the problem \eqref{eq:prob}; and (\romannum{3}) ensure the stability of the original closed-loop system by stabilizing the PCE-transformed system.

\section{Preliminaries on polynomial chaos theory}\label{sect:pce}
This section is a brief overview on polynomial chaos theory to facilitate the PCE-based control synthesis proposed in the next sections. 

For a random vector $\boldsymbol{\xi}$, a function $\psi(\boldsymbol{\xi}): \mathbb{R}^{n_\xi} \rightarrow \mathbb{R}$ with a finite second-order moment admits a PCE \cite{Xiu2002}
\begin{equation*}\label{eq:pce_infty}
\psi(\boldsymbol{\xi}) = \sum\limits_{i=0}^\infty \psi_i \phi_i(\boldsymbol{\xi}),
\end{equation*}
where $\{\psi_i\}$ denotes the expansion coefficients and $\{\phi_i(\boldsymbol{\xi})\}$ denotes the multivariate PC bases. 
By using the Askey scheme of orthogonal polynomial bases, this expansion exponentially converges in the $\mathcal{L}_2$ sense, which results in accurate approximations even with a relatively small number of terms \cite{Xiu2002}.
For the sake of notation simplicity, these basis functions are normalized in the rest of this paper such that they become orthonormal with respect to the probabilistic distribution $\mu (\boldsymbol{\xi})$ of the random vector $\boldsymbol{\xi}$, i.e.,
\begin{align}\label{eqn:innerproduct}
\left\langle \phi_i(\boldsymbol{\xi}),\phi_j(\boldsymbol{\xi}) \right\rangle &= \int_\Xi \phi_i(\boldsymbol{\xi})\phi_j(\boldsymbol{\xi}) \mu(\boldsymbol{\xi}) \ud \boldsymbol{\xi} = \mathbb{E}_{\boldsymbol{\xi}} \{ \phi_i(\boldsymbol{\xi}) \phi_j(\boldsymbol{\xi}) \}\nonumber \\
&= 
\begin{cases}
1 & \text{if } i=j\\
0 & \text{otherwise},
\end{cases}
\end{align}
where $\Xi$ is the support of $\mu(\boldsymbol{\xi})$. 
By exploiting the orthonormality in \eqref{eqn:innerproduct}, each PCE coefficient $\psi_i$ is computed by
\begin{equation}\label{eq:proj_coeff}
\psi_i = \langle \psi(\boldxi), \phi_i(\boldxi)  \rangle,
\end{equation}
which can be calculated via numerical integration \cite{Xiu2002}.
In particular, the mean and variance of $\psi(\boldxi)$ can be computed from the PCE coefficients as below:
\begin{equation}\label{eq:mean_var}
\begin{aligned}
\mathbb{E}_{\boldxi}\{ \psi(\boldxi) \} &= \langle \psi(\boldxi), \phi_0(\boldxi)  \rangle = \psi_0, \\
\text{var}\{ \psi(\boldxi) \} &= \sum_{i=1}^{\infty} (\psi_i \phi_i(\boldxi))^2 = \sum_{i=1}^{\infty} \psi_i^2.
\end{aligned}
\end{equation} 

In practical computations, a PCE with an infinite number of terms is truncated to a finite degree $p$:
\begin{equation}\label{eq:pce_truncated}
\psi(\boldsymbol{\xi}) \approx \hat \psi(\boldsymbol{\xi}) = \sum\limits_{i=0}^{N_p} \psi_i \phi_i(\boldsymbol{\xi}) .
\end{equation}
The total number of terms in (\ref{eq:pce_truncated}) is $N_p+1 = \frac{(n_\xi + p)!}{n_\xi ! p !}$,
depending on the dimension $n_\xi$ of $\boldsymbol{\xi}$ and the highest degree $p$ of the retained polynomials $\{ \phi_i(\boldsymbol{\xi}) \}_{i=0}^{N_p}$.

Let $\boldxi_{[S_i]}$ denote a monomial $\xi_1^{s_{i,1}} \xi_2^{s_{i,2}} \cdots \xi_{n_\xi}^{s_{i,n_\xi}}$ with $S_i = (s_{i,1}, s_{i,2}, \cdots, s_{i,n_\xi})$. The monomial has an exact PCE 
\begin{equation}\label{eq:mono_pce}
\boldxi_{[S_i]} = \sum\limits_{k=0}^{N_{p_i}} \beta_{ik} \phi_k(\boldxi)
\end{equation} 
with a degree $p_i=\sum_{j=1}^{n_{\xi}} s_{i,j}$ without any truncation errors,
since any monomial can be exactly expressed as a linear combination of the corresponding orthonormal polynomial bases, as illustrated by the below example.

\begin{exmp}
	The normalized Legendre polynomials form orthonormal bases with respect to a random scalar $\xi$ uniformly distributed over $[-1,1]$. The first four normalized Legendre polynomials are given by 
	\begin{equation*}
	\left[ \begin{matrix}
	\phi_0(\xi) \\
	\phi_1(\xi) \\
	\phi_2(\xi) \\
	\phi_3(\xi)
	\end{matrix} \right]
	= \left[\begin{matrix}
	1 & 0 & 0 & 0 \\
	0 & {\sqrt{3}} & 0 & 0 \\
	-\frac{\sqrt{5}}{2} & 0 & \frac{3\sqrt{5}}{2} & 0 \\
	0 & -\frac{3\sqrt{7}}{2} & 0 & \frac{5\sqrt{7}}{2} \\
	\end{matrix}\right]
	\left[ \begin{matrix}
	1 \\ \xi \\ \xi^2 \\ \xi^3
	\end{matrix} \right],
	\end{equation*}
	such that $\{ \phi_i(\xi), i=0,1,2,3 \}$ are orthonormal,
	see Appendix B of \cite{LeMa2010}.
	By inverting the coefficient matrix, the monomials $\xi$, $\xi^2$, and $\xi^3$ can be represented as
	$$
	\left[ \begin{matrix}
	\xi \\ \xi^2 \\ \xi^3
	\end{matrix} \right] 
	= \left[\begin{matrix}
	0 & \frac{1}{\sqrt{3}} & 0 & 0 \\
	\frac{1}{{3}} & 0 & \frac{2}{3\sqrt{5}} & 0 \\
	0 & \frac{\sqrt{3}}{5} & 0 & \frac{2}{5\sqrt{7}} \\
	\end{matrix}\right]
	\left[ \begin{matrix}
	\phi_0(\xi) \\
	\phi_1(\xi) \\
	\phi_2(\xi) \\
	\phi_3(\xi)
	\end{matrix} \right],
	$$	
where each row of the above coefficient matrix is the PCE coefficients for the corresponding monomial.
\end{exmp}

With the monomial bases $\{ \boldxi_{[S_i]} \}$, any multivariate polynomial function $\psi(\boldxi)$ is expressed as
\begin{equation}\label{eq:psi_poly}
\psi(\boldxi) = \sum\limits_{S_i \in \mathcal{S}} \alpha_{[S_i]} \boldxi_{[S_i]},
\end{equation}
where $\{ \alpha_{[S_i]} \}$ are the monomial coefficients, and the set $\mathcal{S}$ is determined by all the monomials presented in $\psi(\boldxi)$.
Then the exact PCE for $\psi(\boldxi)$ in \eqref{eq:psi_poly}
can be directly computed from \eqref{eq:mono_pce} as
\begin{equation}\label{eq:poly_pce}
\psi(\boldxi) = \sum_{k=0}^{N_p}\left(\sum_{S_i \in \mathcal{S}} \alpha_{[S_i]} \beta_{ik} \right) \phi_k(\boldxi)
\end{equation}
with $p=\max_{i}\{p_i \}$ denoting the degree of $\psi(\boldxi)$ in \eqref{eq:psi_poly}.

\section{PCE-transformed closed-loop system}\label{sect:pce_cldyn}
In this section, a PCE-transformed system is constructed for the closed-loop system \eqref{eq:clsys}, which will be used in PCE-based control synthesis.

Let $x_i$ denote the $i^\text{th}$ component of the state vector $\mathbf{x}$.
The scalar $x_i(t,\boldsymbol{\xi})$ is expressed as
\begin{align}
x_i(t,\boldsymbol{\xi}) &= \hat x_{i}(t,\boldsymbol{\xi}) + \tilde x_{i}(t,\boldsymbol{\xi}), \nonumber \\
\hat x_{i}(t,\boldsymbol{\xi}) &= \sum_{j=0}^{N_p} x_{i,j}(t)\phi_j(\boldsymbol{\xi}) = \mathbf{x}_i^\top (t) \boldsymbol{\phi}(\boldxi),
\end{align}
where $\hat x_{i}(t,\boldsymbol{\xi})$ is the truncated PCE with a degree $p$, 
$\mathbf{x}_i^\top (t)$ and $\boldsymbol{\phi}(\boldxi)$ denote
\begin{align}
\mathbf{x}^\top_i(t) & = 
\left[ \begin{matrix}
x_{i,0}(t) & x_{i,1}(t) & \cdots & x_{i,N_p}(t)
\end{matrix} \right], \nonumber\\
\boldsymbol{\phi}(\boldxi) & = 
\left[ \begin{matrix}
\phi_0(\boldsymbol{\xi}) &
\phi_1(\boldsymbol{\xi}) & \cdots &
\phi_{N_p}(\boldsymbol{\xi})
\end{matrix} \right]^\top, \label{eq:phi} 
\end{align}
and $\tilde x_{i}(t,\boldsymbol{\xi})$ represents the truncation error.
Define 
\begin{align}
\mathbf{\hat x}(t,\boldsymbol{\xi}) &= 
\left[ \begin{matrix}
\hat x_1(t,\boldsymbol{\xi}) & \hat x_2(t,\boldsymbol{\xi}) & \cdots &
\hat x_{n_x}(t,\boldsymbol{\xi}) 
\end{matrix} \right]^\top, \nonumber \\
\mathbf{\tilde x}(t,\boldsymbol{\xi}) &= 
\left[ \begin{matrix}
\tilde x_1(t,\boldsymbol{\xi}) & \tilde x_2(t,\boldsymbol{\xi}) & \cdots &
\tilde x_{n_x}(t,\boldsymbol{\xi}) 
\end{matrix} \right]^\top, \nonumber \\
\mathbf{x}_{\mathrm{PCE}}(t) &= \begin{bmatrix}
\mathbf{x}_1(t) & \cdots & \mathbf{x}_{n_x}(t)
\end{bmatrix}, \nonumber
\end{align}
then the PCE of the vector $\mathbf{x}(t,\boldsymbol{\xi})$ can be written as 
\begin{equation}\label{eq:vecs_pce}
\begin{aligned}
\mathbf{x}(t,\boldsymbol{\xi}) &= \mathbf{\hat x}(t,\boldsymbol{\xi}) + \mathbf{\tilde x}(t,\boldsymbol{\xi}) = \mathbf{x}_{\mathrm{PCE}}^\top(t)
\boldsymbol{\phi}(\boldsymbol{\xi}) 
+ \mathbf{\tilde x}(t,\boldsymbol{\xi})\\
&= \underbrace{\left( \boldsymbol{\phi}^\top (\boldxi) \otimes I_{n_x} \right)}_{\Phi_\mathbf{x}^\top(\boldxi)} 
\underbrace{\text{vec}\left( \mathbf{x}_{\mathrm{PCE}}^\top(t) \right)}_{\mathbf{X}(t)}
+ \mathbf{\tilde x}(t,\boldsymbol{\xi}).
\end{aligned}
\end{equation}
The last equation of \eqref{eq:vecs_pce} applies
the property of the Kronecker product $\text{vec}(EFG) = (G^\top \otimes E) \text{vec}(F)$.

With the PCE of $\mathbf{x}(t,\boldsymbol{\xi})$ in the form of (\ref{eq:vecs_pce}), the closed-loop system (\ref{eq:clsys}) can be equivalently rewritten as
\begin{subequations}\label{eq:pce_insert_dynall}
\begin{align}
\Phi_\mathbf{x}^\top(\boldxi) \mathbf{\dot X} (t) &= \mathbf{A}_\text{cl}(\boldxi)\Phi_\mathbf{x}^\top(\boldxi)\mathbf{X}(t) + \mathbf{B}_{\text{cl}}(\boldxi)
\mathbf{w}(t) + \mathbf{r}_\mathbf{x}(t,\boldsymbol{\xi}), 
\label{eq:pce_insert_dyn}\\
\mathbf{r}_\mathbf{x} (t,\boldsymbol{\xi}) &= - \mathbf{\dot {\tilde x}}(t,\boldsymbol{\xi}) + \mathbf{A}_\text{cl}(\boldxi) \mathbf{\tilde x}(t,\boldxi), \label{eq:rx} \\
\mathbf{z}(t,\boldxi)  & = \mathbf{z}_\text{nom}(t,\boldxi) + \mathbf{r}_\mathbf{z}(t,\boldxi), \label{eq:pce_insert_out} \\
\mathbf{z}_\text{nom}(t,\boldxi) & = \mathbf{C}_{\text{cl}}(\boldxi)\Phi_\mathbf{x}^\top(\boldxi)\mathbf{X}(t) + \mathbf{D}_\text{cl}(\boldxi)\mathbf{w}(t), \label{eq:znom} \\
\mathbf{r}_\mathbf{z}(t,\boldxi) &=
\mathbf{C}_{\text{cl}}(\boldxi) \mathbf{\tilde x}(t,\boldxi). \label{eq:rz}
\end{align}
\end{subequations}
The PCE-based transformation of of the closed-loop dynamic equation \eqref{eq:clsys_dyn} follows the typical stochastic Galerkin method \eqref{eq:pce_insert_dynall} as follows. 
Left-multiplying (\ref{eq:pce_insert_dyn}) by $\Phi_\mathbf{x}(\boldxi)$ leads to
\begin{equation}\label{eqn:transformationstep2}
\begin{aligned}
& \Phi_\mathbf{x}(\boldxi)\Phi_\mathbf{x}^\top(\boldxi)
\mathbf{\dot X}(t) = \Phi_\mathbf{x}(\boldxi)\mathbf{A}_\text{cl}(\boldxi)\Phi_\mathbf{x}^\top(\boldxi)
\mathbf{X}(t) \\
&\qquad\qquad\qquad + \Phi_\mathbf{x}(\boldxi)
\mathbf{B}_\text{cl}(\boldxi)
\mathbf{w}(t) 
+ \Phi_\mathbf{x}(\boldxi) \mathbf{r}_\mathbf{x}(t, \boldxi).
\end{aligned}
\end{equation}
Taking the mathematical expectation with respect to $\boldsymbol{\xi}$ on both sides of (\ref{eqn:transformationstep2}), the PCE-transformed closed-loop dynamic equation is obtained as
\begin{subequations}\label{eq:pce_dyn_all}
\begin{align}
\dot{\mathbf{X}}(t) & = \mathcal{\bar A}_\text{cl}\mathbf{X}(t) + \mathcal{\bar B}_\text{cl}\mathbf{w}(t) + \mathbf{\bar R}_\mathbf{x}(t), \label{eq:pce_dyn}\\
\mathbf{\bar R}_\mathbf{x}(t) &= \mathbb{E}_{\boldxi}
\{\Phi_\mathbf{x}(\boldxi)
\mathbf{r}_\mathbf{x}(t, \boldxi)\}
= \mathbb{E}_{\boldxi}
\{\Phi_\mathbf{x}(\boldxi)
\mathbf{A}_\text{cl}(\boldxi) \mathbf{\tilde x}(t,\boldxi)\}, \label{eqn:BRx} \\
\mathcal{\bar A}_\text{cl} &= \mathbb{E}_{\boldxi}\{ \Phi_\mathbf{x}(\boldxi) \mathbf{A}_\text{cl}(\boldxi) \Phi_\mathbf{x}^\top(\boldxi) \}, \label{eq:barAcl}\\
\mathcal{\bar B}_\text{cl} &= \mathbb{E}_{\boldxi}\{ \Phi_\mathbf{x}(\boldxi) \mathbf{B}_{\text{cl}}(\boldxi) \}, \label{eq:barBcl}
\end{align}
\end{subequations}
which describes the dynamics of the PCE coefficient vector $\mathbf{X}(t)$.
The above derivations exploit  $\mathbb{E}_{\boldxi}\{\Phi_\mathbf{x}(\boldxi)\Phi_\mathbf{x}^\top(\boldxi)\} =  \mathbf{I}_{n_x(N_p+1)}$ and 
$\mathbb{E}_{\boldxi}\{\Phi_\mathbf{x}(\boldxi) \mathbf{\dot {\tilde x}}(t,\boldxi)\} = \mathbf{0}$, according to (\ref{eqn:innerproduct}) and the definition of $\Phi_\mathbf{x}(\boldxi)$ in \eqref{eq:phi} and \eqref{eq:vecs_pce}. 

To compute $\mathcal{\bar A}_\text{cl}$ and 
$\mathcal{\bar B}_\text{cl}$ defined in \eqref{eq:barAcl} and \eqref{eq:barBcl}, the polynomial dependence of $\Phi_\mathbf{x}(\boldxi) \mathbf{B}(\boldxi)$, $\mathbf{C}(\boldxi)\Phi_\mathbf{x}^\top (\boldxi)$, and $\mathbf{D}_\mathbf{w}(\boldxi)$ on $\boldxi$ should be exploited according to Assumption \ref{ass:polymat}.
Let $q$ denote the maximal degree of these polynomial matrices above. 
Then according to \eqref{eq:mono_pce}--\eqref{eq:poly_pce}, their exact PCEs are expressed as 
\begin{subequations}\label{eq:phi_BCDw}
	\begin{align}
	\Phi_\mathbf{x}(\boldxi) \mathbf{B}(\boldxi) &= \sum_{i=0}^{N_q} \mathbf{\hat B}_i \phi_i(\boldxi), \label{eq:hatBi}\\
	\mathbf{C}(\boldxi)\Phi_\mathbf{x}^\top (\boldxi) &= \sum_{i=0}^{N_q} \mathbf{\hat C}_i \phi_i(\boldxi), \label{eq:hatCi}\\
	\mathbf{D}_\mathbf{w}(\boldxi) &= \sum_{i=0}^{N_q}
	\mathbf{\hat D}_{\mathbf{w},i} \phi_i(\boldxi), \label{eq:hatDwi}
	\end{align}
\end{subequations}
where $N_q+1$ is the number of terms in the PCE of degree $q$.

\begin{proposition}\label{prop:Acl_Bcl}
With the PCE coefficients $\{\mathbf{\hat B}_i \}$, $\{\mathbf{\hat C}_i \}$, and $\{\mathbf{\hat D}_{\mathbf{w},i} \}$ in \eqref{eq:hatBi}--\eqref{eq:hatDwi},
$\mathcal{\bar A}_\text{cl}$ in \eqref{eq:barAcl} and 
$\mathcal{\bar B}_\text{cl} $ in \eqref{eq:barBcl} are computed as 
\begin{subequations}\label{eq:ABcl}
\begin{align}
\mathcal{\bar A}_\text{cl} &= \mathcal{A} + \mathcal{\bar B}\mathcal{\bar K}\mathcal{\bar C}, \;
\mathcal{\bar B}_\text{cl} = \mathcal{B}_\mathbf{w} + \mathcal{\bar B}\mathcal{\bar K}\mathcal{\bar D}_\mathbf{w}, \label{eq:Acl_Bcl_pc}\\
\mathcal{A} &= \mathbb{E}_{\boldxi}\{\Phi_\mathbf{x}(\boldxi) \mathbf{A}(\boldxi) \Phi_\mathbf{x}^\top(\boldxi)\}, \label{eq:A_pc}\\
\mathcal{B}_\mathbf{w} &=   \mathbb{E}_{\boldxi}
\{\Phi_\mathbf{x}(\boldxi)
\mathbf{B}_\mathbf{w}(\boldxi)\}, \label{eq:Bw_pc}\\
\mathcal{\bar B} &= \begin{bmatrix}
\mathbf{\hat B}_0 & \mathbf{\hat B}_1 & \cdots & \mathbf{\hat B}_{N_q} 
\end{bmatrix}\!, \label{eq:barB_pc}\\
\mathcal{\bar C} &= \begin{bmatrix}
\mathbf{\hat C}_0^\top & \mathbf{\hat C}_1^\top & \cdots & \mathbf{\hat C}_{N_q}^\top 
\end{bmatrix}^\top\!, \label{eq:hatC_pc}\\
\mathcal{\bar D}_\mathbf{w} &= \begin{bmatrix}
\mathbf{\hat D}_{\mathbf{w},0}^\top & \mathbf{\hat D}_{\mathbf{w},1}^\top & \cdots & \mathbf{\hat D}_{\mathbf{w},N_q}^\top 
\end{bmatrix}^\top\!, \label{eq:barDw_pc}\\
\mathcal{\bar K} &= \mathbf{I}_{N_q+1} \otimes \mathbf{K}. \label{eq:hatK_pc}
\end{align}
\end{subequations}
\end{proposition}

The proof of Proposition \ref{prop:Acl_Bcl} is in Appendix \ref{app:Acl_Bcl_pc}.

The matrices $\mathcal{A}$ in \eqref{eq:A_pc}, $\mathcal{B}_\mathbf{w}$ in \eqref{eq:Bw_pc}, 
$\{\mathbf{\hat B}_i \}$ in \eqref{eq:hatBi}, $\{\mathbf{\hat C}_i \}$ in \eqref{eq:hatCi}, and $\{\mathbf{\hat D}_{\mathbf{w},i} \}$ in \eqref{eq:hatDwi} can be computed by either projections as in \eqref{eq:proj_coeff}--\eqref{eq:mean_var} via numerical integration or the use of exact PCEs of monomials as in \eqref{eq:mono_pce}--\eqref{eq:poly_pce}.

The PCE-transformed approximation of the output equation \eqref{eq:pce_insert_out} is described in the proposition below, which is different from the standard Galerkin projection.

\begin{proposition}\label{prop:z}
The equation
\begin{equation}\label{eq:Znom_eq}
\| \mathbf{Z}_\text{nom}(t) \|_{\mathcal{L}_2}^2 = \mathbb{E}_{\boldxi}\{ \left\| \mathbf{z}_\text{nom}(t,\boldxi) \right\|_{\mathcal{L}_2}^2 \}
\end{equation}
holds, where
$\mathbf{z}_{\text{nom}}(t,\boldxi)$ is defined in \eqref{eq:znom}, and 
$\mathbf{Z}_\text{nom}(t)$ is defined by
\begin{subequations}\label{eq:CDcl}
\begin{align}
\mathbf{Z}_\text{nom}(t)
&= \mathcal{\bar C}_\text{cl} \mathbf{X}(t) + \mathcal{\bar D}_\text{cl} \mathbf{w}(t),  \label{eq:zhat_2norm}\\
\mathcal{\bar C}_\text{cl} &= \mathcal{\bar C}_\mathbf{Z} + \mathcal{\bar D}_\mathbf{Z} \mathcal{\bar K} \mathcal{\bar C}, \;
\mathcal{\bar D}_\text{cl} = \mathcal{\bar D}_\mathbf{Zw} + \mathcal{\bar D}_\mathbf{Z} \mathcal{\bar K} \mathcal{\bar D}_\mathbf{w}, \label{eq:Ccl_Dcl}\\
\mathcal{\bar C}_\mathbf{Z} &= \begin{bmatrix}
\mathcal{C}_\mathbf{Z} \\ \mathbf{0}
\end{bmatrix} \!\in \mathbb{R}^{n_z(N_q+1) \times n_x(N_p+1)}, \label{eq:barCZ_pc}\\
\mathcal{\bar D}_\mathbf{Zw} &= \begin{bmatrix}
\mathcal{D}_\mathbf{Zw} \\ \mathbf{0}
\end{bmatrix} \!\in \mathbb{R}^{n_z(N_q+1) \times n_w}, \\
\mathcal{C}_\mathbf{Z} &= \mathbf{I}_{N_p+1} \otimes \mathbf{C}_\mathbf{z}, \;
\mathcal{D}_\mathbf{Zw} = \mathbb{E}_{\boldxi}\{ \Phi_\mathbf{z}(\boldxi) \mathbf{D}_\mathbf{zw} \}, \label{eq:Cz_Dzw_pc} \\
\mathcal{\bar D}_\mathbf{Z} &= \mathbf{I}_{N_q+1} \otimes \mathbf{D}_\mathbf{z}, \label{eq:hatDz_pc}
\end{align}
\end{subequations}
with $\Phi_\mathbf{z}(\boldxi)$ defined similarly to $\Phi_\mathbf{x}(\boldxi)$ in \eqref{eq:vecs_pce}.
\end{proposition}

The proof of Proposition \ref{prop:z} is in Appendix \ref{app:z}.

The above derivations \eqref{eq:pce_insert_dynall}--\eqref{eq:pce_dyn_all} are exact, including the effect of unknown truncation errors $\mathbf{\tilde x}(t,\boldxi)$ captured by $\mathbf{r_x}(t,\boldxi)$, $\mathbf{r_z}(t,\boldxi)$, and $\mathbf{\bar R_x}(t)$. After neglecting these unknown terms, a PCE-transformed approximation 
\begin{equation}\label{eq:Xa}
	\begin{aligned}
		\dot{\mathbf{X}}_a (t) & = \mathcal{\bar A}_\text{cl}\mathbf{X}_a(t) + \mathcal{\bar B}_\text{cl}\mathbf{w}(t) \\
		\mathbf{Z}_a(t)
		&= \mathcal{\bar C}_\text{cl} \mathbf{X}_a (t) + \mathcal{\bar D}_\text{cl} \mathbf{w}(t),
	\end{aligned}
\end{equation}
is obtained according to Propositions \ref{prop:Acl_Bcl} and \ref{prop:z}.
Note that $\mathbf{X}_a(t)$ and $\mathbf{Z}_a(t)$ in \eqref{eq:Xa} are approximates of $\mathbf{X}(t)$ in \eqref{eq:pce_dyn_all} and $\mathbf{Z}_\text{nom}(t)$ in \eqref{eq:zhat_2norm}.
This PCE-transformed system \eqref{eq:Xa} is different from the existing ones in two aspects:
\begin{enumerate}
	\item[(\romannum{1})] The existing approach in \cite{Fisher2009,Shen2017,Wan2018ACC} 
	separately transforms the four system equations in \eqref{eqn:open_sys} and \eqref{eqn:sofcontroller} by applying standard Galerkin projection, to construct a PCE-transformed closed-loop system.
	In this case, if the open-loop system is unstable due to its parametric uncertainties, the divergent open-loop system state as a function of $\boldsymbol{\xi}$ would not admit a PCE. In contrast, our proposed transformation directly considers the original closed-loop dynamics, which allows PCEs of system states by stabilizing the closed-loop dynamics.
	\item[(\romannum{2})] With the error term $\mathbf{\bar R_x}(t)$ removed from \eqref{eq:pce_dyn}, the squared $\mathcal{L}_2$-induced gain from $\mathbf{w}(t)$ to $\mathbf{z}_\text{nom}(t,\boldxi)$ in \eqref{eq:pce_dyn} and \eqref{eq:znom} is equal to the squared $\mathcal{H}_\infty$-norm of the PCE-transformed approximation \eqref{eq:Xa}, i.e.,
	\begin{equation}\label{eq:Hinf_eq}
		\sup\limits_{ \left\| \mathbf{w} \right\|_{\mathcal{L}_2}^2 \leq 1 } \mathbb{E}_{\boldxi} \left\{ \left\| \mathbf{z}_\mathrm{nom}(t,\boldxi) \right\|_{\mathcal{L}_2}^2 \right\} =
		\sup\limits_{ \left\| \mathbf{w} \right\|_{\mathcal{L}_2}^2 \leq 1 } \left\| \mathbf{Z}_\text{nom}(t) \right\|_{\mathcal{L}_2}^2,
	\end{equation}
	according to \eqref{eq:Znom_eq}. This property does not hold for the PCE-transformed approximation given in \cite{Fisher2009,Shen2017,Wan2018ACC}. 
\end{enumerate} 

Moreover, it is proved by \eqref{eq:app_err}--\eqref{eq:up_bound} in Appendix \ref{app:comp} that the worst-case upper bound for the approximation error $\mathbf{x}(t,\boldxi) - \Phi_\mathbf{x}^\top (\boldxi) \mathbf{X}_a(t)$ from \eqref{eq:Xa} is smaller than the one obtained from 
the PCE-transformed system in \cite{Fisher2009,Shen2017,Wan2018ACC}.

\section{Robust PCE-based Static Output-Feedback Control}\label{sect:sof_synthesis}
The original synthesis problem \eqref{eq:prob} can be transformed into a standard $\mathcal{H}_{\infty}$-control problem 
\begin{equation*}\label{eq:prob1}
		\begin{aligned}
			\min_{\mathbf{K},\gamma_\text{nom}>0}& \; \gamma_\text{nom}  \\
			\text{s.t.}&\; \left\| \mathbf{Z}_\text{nom}(t) \right\|_{\mathcal{L}_2}^2 
			\leq \gamma_\text{nom}^2 \left\| \mathbf{w}(t) \right\|_{\mathcal{L}_2}^2, 
		\end{aligned}
\end{equation*}
for the PCE-transformed closed-loop system \eqref{eq:Xa}.
By applying the bounded real lemma \cite{dullerud2013course}, the above problem is reformulated as the following synthesis problem \cite{Wan2018ACC}:
\begin{equation}\label{eq:prob1_mi}
		\begin{aligned}
			&\min_{\mathbf{P},\mathbf{K},\gamma_\text{nom}} \; \gamma_\text{nom} \\ 
			&\text{s.t.} \; 
			\mathbf{P}>0,\, \gamma_\text{nom}>0, \\
			&\quad \begin{bmatrix}
				\text{He}\{\mathbf{P} \mathcal{\bar A}_\text{cl}\}
				& \mathbf{P} \mathcal{\bar B}_{\text{cl}} & \mathcal{\bar C}_{\text{cl}}^\top \\
				\mathcal{\bar B}_{\text{cl}}^\top \mathbf{P} & 
				-\gamma_\text{nom} \mathbf{I}_{n_w} & \mathcal{\bar D}_{\text{cl}}^\top \\
				\mathcal{\bar C}_{\text{cl}} & \mathcal{\bar D}_{\text{cl}} & -\gamma_\text{nom} \mathbf{I}_{n_x(N_p+1)}
			\end{bmatrix} < 0.
		\end{aligned}
\end{equation}
The third constraint in (\ref{eq:prob1_mi}) is a bilinear matrix inequality (BMI) due to the multiplication between $\mathbf{P}$ and $\mathcal{\bar K}$ in $\mathbf{P} \mathcal{\bar A}_\text{cl}$ and $\mathbf{P} \mathcal{\bar B}_{\text{cl}}$. These bilinear terms cannot be converted to linear terms by using the conventional change of variables, due to the block-diagonal structure of $\mathcal{\bar K} = \mathbf{I}_{N_q+1} \otimes \mathbf{K}$.
	
It should be noted that the PCE truncation error is neglected in \eqref{eq:Xa}, thus is not addressed by the above synthesis problem \eqref{eq:prob1_mi}. Due to this reason, even if the nominal PCE-transformed closed-loop system \eqref{eq:Xa} is stabilized, the true closed-loop system \eqref{eq:clsys} is not necessarily stable. 
A commonly adopted remedy in literature is to use higher-degree PCEs with smaller truncation errors, which still provides no theoretical guarantee. In addition, as the PCE degree $p$ becomes higher, the number of PCE terms grows factorially, which significantly increases the computational complexity involved in solving the PCE-based synthesis problem. 
	
To address the above stability issue, a robust PCE-based control synthesis method is proposed in this section by constructing a norm-bounded LDI to describe the effect of PCE truncation errors. This approach robustifies stability and performance without increasing the PCE degree. 

\subsection{Linear differential inclusion for PCE-transformed closed-loop system}\label{sect:pce_transf}
Before deriving a LDI for the PCE-transformed closed-loop system, the following proposition is given for the PCE truncation error $\mathbf{\tilde x}(t,\boldxi)$.

\begin{proposition}\label{prop:MN}
For the PCE truncation error $\mathbf{\tilde x}(t,\boldxi)$, there exists a non-unique matrix $\mathbf{M}(t,\boldxi) \in \mathbb{R}^{n_x \times n_x}$ such that 
\begin{equation}\label{eq:xtilde2xpce}
	\mathbf{\tilde x}(t,\boldxi) = \mathbf{M}(t,\boldxi) \Phi_\mathbf{x}^\top (\boldxi)  \mathbf{X}(t)
	= \Phi_\mathbf{x}^\top (\boldxi) \mathbf{N}(t,\boldxi) \mathbf{X}(t),
\end{equation}
with $\mathbf{N}(t,\boldxi)$ defined as 
$\mathbf{N}(t,\boldxi) = \mathbf{I}_{N_p+1} \otimes \mathbf{M}(t,\boldxi)$.
\end{proposition}

The proof of Proposition \ref{prop:MN} is given in Appendix \ref{app:MN}.

It is assumed in the following derivations that the uncertainty matrix $\mathbf{N}(t,\boldxi)$ in \eqref{eq:xtilde2xpce} is norm-bounded, i.e., 
\begin{equation}\label{eq:Fx}
\mathbf{N}(t,\boldxi) \in \mathcal{F}_\mathbf{x} = \{ 
\boldsymbol{\Delta}  \;|\; \boldsymbol{\Delta}^\top \boldsymbol{\Delta} \leq \rho^2 \mathbf{I}_{n_x(N_p+1)}  \}.
\end{equation}
Then, the norm-bounded LDI system \cite{dullerud2013course}
\begin{subequations}\label{eq:pce_sys_LDI}
	\begin{align}
	\Phi_\mathbf{x}^\top(\boldxi) \mathbf{\dot X}_1 (t) &= \mathbf{A}_\text{cl}(\boldxi)\Phi_\mathbf{x}^\top(\boldxi)
	(\mathbf{I} + \boldsymbol{\Delta}_\mathbf{x}(t))
	\mathbf{X}_1(t)  \nonumber \\
	&\quad + \mathbf{B}_{\text{cl}}(\boldxi)
	\mathbf{w}(t) - \mathbf{\dot {\tilde x}}(t,\boldsymbol{\xi})
	\label{eq:pce_sys_LDI_dyn}\\
	\mathbf{z}_\text{rob}(t,\boldxi) &= \mathbf{C}_{\text{cl}}(\boldxi)\Phi_\mathbf{x}^\top(\boldxi)
	(\mathbf{I} + \boldsymbol{\Delta}_\mathbf{x}(t))
	\mathbf{X}_1(t)  \nonumber \\
	&\quad + \mathbf{D}_\text{cl}(\boldxi)\mathbf{w}(t)
	\label{eq:pce_sys_LDI_out}
	\end{align}
\end{subequations}
is constructed for the closed-loop system \eqref{eq:pce_insert_dynall} by substituting \eqref{eq:xtilde2xpce} into \eqref{eq:rx} and \eqref{eq:rz} and replacing $\mathbf{N}(t,\boldxi)$ with $\boldsymbol{\Delta}_\mathbf{x}(t) \in \mathcal{F}_\mathbf{x}$, where $\mathbf{I}$ represents an identity matrix of appropriate dimensions.
Under the condition \eqref{eq:xtilde2xpce}, 
the system trajectory set of the LDI \eqref{eq:Fx}--\eqref{eq:pce_sys_LDI} includes the system trajectory of \eqref{eq:pce_insert_dynall}.
Hence, given $\mathbf{X}(0)=\mathbf{X}_1(0)=0$, the following inequality holds:
\begin{equation}\label{eq:znorm_ineq}
\begin{aligned}
\sup_{\mathbf{w}(t),\, \boldsymbol{\Delta}_\mathbf{x}(t) \in \mathcal{F}_\mathbf{x}} \;
\mathbb{E}_{\boldxi}\{ \left\| \mathbf{z}_\text{rob}(t,\boldxi) \right\|_{\mathcal{L}_2}^2 \}
\geq \mathbb{E}_{\boldxi}\{ \left\| \mathbf{z}(t,\boldxi) \right\|_{\mathcal{L}_2}^2 \}.
\end{aligned}
\end{equation} 

With similar procedures as in Section \ref{sect:pce_cldyn}, the PCE-transformed LDI, i.e.,
the PCE transformation of \eqref{eq:pce_sys_LDI}, is obtained as
\begin{subequations}\label{eq:pce_sys_unc}
\begin{align}
\dot{\mathbf{X}}_1(t) & = \mathcal{\bar A}_\text{cl} (\mathbf{I} + \boldsymbol{\Delta}_\mathbf{x}(t)) \mathbf{X}_1(t) + \mathcal{\bar B}_\text{cl}\mathbf{w}(t),  \label{eq:pce_sys_unc_dyn}\\
\mathbf{Z}_\text{rob}(t) &= \mathcal{\bar C}_{\text{cl}}
(\mathbf{I} + \boldsymbol{\Delta}_\mathbf{x}(t))
\mathbf{X}_1(t) + \mathcal{\bar D}_\text{cl}\mathbf{w}(t), 
\label{eq:pce_sys_unc_out}
\end{align}
\end{subequations}
where $\mathcal{\bar A}_\text{cl}$, $\mathcal{\bar B}_\text{cl}$, $\mathcal{\bar C}_\text{cl}$, and $\mathcal{\bar D}_\text{cl}$
are computed according to Propositions \ref{prop:Acl_Bcl} and \ref{prop:z}. Similarly to \eqref{eq:Znom_eq}, the above PCE transformation ensures
$\| \mathbf{Z}_\text{rob}(t) \|_{\mathcal{L}_2}^2 = \mathbb{E}_{\boldxi}\{ \left\| \mathbf{z}_\text{rob}(t,\boldxi) \right\|_{\mathcal{L}_2}^2 \}$, which further implies 
\begin{equation}\label{eq:znorm_ineq2}
\begin{aligned}
\sup_{\mathbf{w}(t), \, \boldsymbol{\Delta}_\mathbf{x}(t) \in \mathcal{F}_\mathbf{x}} \;
 \left\| \mathbf{Z}_\text{rob}(t) \right\|_{\mathcal{L}_2}^2 
\geq \mathbb{E}_{\boldxi}\{ \left\| \mathbf{z}(t,\boldxi) \right\|_{\mathcal{L}_2}^2\}
\end{aligned}
\end{equation}
for any $\mathbf{X}(0)=\mathbf{X}_1(0)=0$ according to \eqref{eq:znorm_ineq}.

\subsection{Robust PCE-based control synthesis}\label{sect:rb_pce_syn}

Based on the PCE-transformed LDI \eqref{eq:pce_sys_unc}, a $\mathcal{H}_\infty$ control synthesis problem can be formulated as 
\begin{equation}\label{eq:prob2}
\begin{aligned}
\min_{\mathbf{K},\gamma_\text{rob}>0}& \quad \gamma_\text{rob}  \\
\text{s.t.}&\quad \left\| \mathbf{Z}_\text{rob}(t) \right\|_{\mathcal{L}_2}^2 
\leq \gamma_\text{rob}^2 \left\| \mathbf{w}(t) \right\|_{\mathcal{L}_2}^2, \\
&\quad \forall \left\| \mathbf{w}(t) \right\|_{\mathcal{L}_2} < \infty, \; 
{\Delta}_\mathbf{x}(t) \in \mathcal{F}_\mathbf{x}.
\end{aligned}
\end{equation}
According to \eqref{eq:znorm_ineq2}, this problem minimizes an upper bound of $\gamma$ in the original problem \eqref{eq:prob}. 

\begin{theorem}\label{thm:rb_pce}
The PCE-transformed LDI \eqref{eq:pce_sys_unc} 
is quadratically stable, and its $\mathcal{H}_\infty$ norm 
is upper bounded by $\gamma_\text{rob}$, if there exist a positive definite matrix $\mathbf{P}$ and a scalar $\tau>0$ such that 
\begin{equation}\label{eq:lmi_syn}
\begin{bmatrix}
\mathrm{He}\{ \mathbf{P} \mathcal{\bar A}_\text{cl}  \}
+ \tau \rho^2 \mathbf{I} & \mathbf{P}\mathcal{\bar B}_{\text{cl}}
& \mathbf{P} \mathcal{\bar A}_\text{cl}
& \mathcal{\bar C}_{\text{cl}}^\top \\
\mathcal{\bar B}_{\text{cl}}^\top \mathbf{P} &
- \gamma_\text{rob} \mathbf{I} & \mathbf{0} & \mathcal{\bar D}_{\text{cl}}^\top \\
\mathcal{\bar A}_\text{cl}^\top \mathbf{P} & \mathbf{0} 
& - \tau \mathbf{I} & \mathcal{\bar C}_{\text{cl}}^\top \\
\mathcal{\bar C}_{\text{cl}}  &
\mathcal{\bar D}_{\text{cl}} & 
\mathcal{\bar C}_{\text{cl}} &
-\gamma_\text{rob} \mathbf{I} 
\end{bmatrix} < 0
\end{equation}
where $\rho^2$ is defined in \eqref{eq:Fx} to quantify the effect of PCE truncation errors.
\end{theorem}

The proof of Theorem \ref{thm:rb_pce} is given in Appendix \ref{app:rb_pce}.

The robust PCE-based control synthesis problem is then formulated as
\begin{equation}\label{eq:prob2_mi}
\begin{aligned}
\min_{\mathbf{P},\mathbf{K},\gamma_\text{rob},\tau} &\quad \gamma_\text{rob} \\ 
\text{s.t.} &\quad \eqref{eq:lmi_syn}, \,
\mathbf{P}>0, \, \gamma_\text{rob} > 0, \, \tau>0.
\end{aligned}
\end{equation}
The constraint \eqref{eq:lmi_syn} is a bilinear matrix inequality (BMI) with respect to $\mathbf{P}$ and $\mathbf{K}$. 
It should be noted that the PCE-based synthesis solution to \eqref{eq:prob2_mi} might fail to stabilize the original closed-loop system \eqref{eq:clsys} if the bound $\rho^2$ in \eqref{eq:Fx} is inadequate to address the underlying uncertainty. This issue brings up the following discussions. 

First of all, the corollary below presents a sufficient condition under which the control gain derived from \eqref{eq:prob2_mi} stabilizes the original closed-loop system \eqref{eq:clsys}. 

\newtheorem{corollary}{Corollary}[theorem]
\begin{corollary}\label{thm:stable}
If the PCE-transformed LDI \eqref{eq:pce_sys_unc} is robustly stabilized and the uncertainty matrix $\mathbf{N}(t,\boldxi)$ in \eqref{eq:xtilde2xpce} remains bounded with probability 1 at any time $t$,
the original closed-loop system \eqref{eq:clsys} is mean-square stable.
\end{corollary}

The proof of the above corollary is given in Appendix \ref{app:cor_stability}.

However, it is a challenging task to determine the uncertainty bound $\rho^2$ in \eqref{eq:Fx} for $\mathbf{N}(t,\boldxi)$. On one hand, $\mathbf{N}(t,\boldxi)$ is related to the PCE truncation error of the \textit{closed-loop} state $\mathbf{x}(t,\boldxi)$, hence it is bounded if the closed-loop state has a convergent PCE for all $t>0$. Since this requirement for the closed-loop state can be check only after constructing the closed-loop system \eqref{eq:clsys}, $\rho^2$ in \eqref{eq:Fx} cannot be verified before solving the synthesis problem \eqref{eq:prob2_mi}.  
On the other hand, it is still challenging to determine the uncertainty bound $\rho^2$ after having a given control gain. Generally, computing $\rho^2$ involves deriving the PCE of $\mathbf{x}(t,\boldxi)$ whose dependence on $\boldxi$ is usually non-polynomial.
This requires numerical integration at each time instant (see Section IV in \cite{Lucia2017}), thus is infeasible to be performed for all $t>0$.

As a remedy to the above difficulty of directly deriving the uncertainty bound $\rho^2$ in \eqref{eq:Fx}, we propose to perform a post-analysis of robust stability and re-tune $\rho^2$ as a robustifying parameter. 
For stability analysis of polynomially uncertain systems considered in this paper, various methods such as polynomially parameter-dependent Lyapunov function \cite{chesi2009homo} and sum-of-squares \cite{chesi2013exact} are available. By combining such stability analysis and an iterative bisection search strategy, a lower bound $\rho_{\min}^2$ can be determined such that any $\rho^2 \geq \rho_{\min}^2$ results in a stabilizing gain for the closed-loop system \eqref{eq:clsys}. The basic idea is as follows.
We start with a value of $\rho^2$ that ensures the feasibility of the synthesis problem \eqref{eq:prob2_mi}. Then, in each iteration, the following successive steps are performed: \romannum{1}) solving the synthesis problem \eqref{eq:prob2_mi}; \romannum{2}) doing the post-analysis of robust stability with the obtained control gain; and \romannum{3}) tuning $\rho^2$ with a bisection search strategy, i.e., decreasing $\rho^2$ if the resulting closed-loop system is stabilized, or increasing $\rho^2$ if it is unstable. 
Re-tuning $\rho^2 \geq \rho_{\min}^2$ involves tradeoffs between stability and performance. The use of a more conservative bound $\rho^2$ enhances stability, but sacrifices control performance. 

\begin{rem}\label{rem:stability}
In \cite{Hsu2020design,Wan2018mixed}, an alternative approach is proposed to ensure the stability of the original closed-loop system \eqref{eq:clsys}. It includes the conventional worst-case stability condition as a complementary constraint for the nominal PCE-based synthesis problem \eqref{eq:prob1_mi}. 
\end{rem}

\begin{rem}\label{rem:diff}
Compared to our previous paper \cite{Wan2018ACC}, the proposed robust PCE-based control synthesis here has the following advantages. 
Firstly, two terms of norm-bounded uncertainties are used in \cite{Wan2018ACC}, while only one term of norm-bounded uncertainty $\Delta_\mathbf{x}(t)$ is introduced in \eqref{eq:pce_sys_LDI}. Hence only one robustifying parameter $\rho$ in \eqref{eq:Fx} needs to be determined, which is much simpler than determining two robustifying parameters in \cite{Wan2018ACC}.
Secondly, the stability issue of the original closed-loop system \eqref{eq:clsys} was just briefly mentioned in \cite{Wan2018ACC}, whilst detailed discussions are given here including Corollary \ref{thm:stable} and post-analysis with the above bisection search strategy to determine $\rho^2$.
\end{rem}

\section{Simulation example}\label{sect:sim}

Consider the system \eqref{eqn:open_sys}:
\begin{equation*}
\mathbf{A}(\xi) = \left[\begin{array}{cc} 0.6\xi^3 & -0.4 \\ 0.1 & 0.5\end{array}\right], \;
\mathbf{B}_\mathbf{w} = \begin{bmatrix}
	1 & 0 & 0 & 0 \\ 
	0 & 1 & 0 & 0
\end{bmatrix}\!,
\end{equation*}
\begin{equation*}\label{eqn:numericalexample}
\begin{aligned}
\mathbf{B}(\xi) &= \begin{bmatrix}
0.2+\xi^3 \\ 0.2
\end{bmatrix}\!,
\mathbf{C}_\mathbf{z} = 
\begin{bmatrix}
1 & 0 & 0 \\
0 & 1 & 0
\end{bmatrix}^\top\!, \\
\mathbf{C}(\xi) &= \begin{bmatrix} 
1 & \xi^3 \\ 0 & 1\end{bmatrix}, \; 
\mathbf{D}_\mathbf{w}(\xi) = \begin{bmatrix}
0 & 0 & 1+2\xi^3 & 0 \\ 
0 & 0 & 0 & 1
\end{bmatrix}\!, \\
\mathbf{D}_\mathbf{z} &= \begin{bmatrix}
0 & 0 & 0.2
\end{bmatrix}^\top\!,
\mathbf{D}_\mathbf{zw} = \mathbf{0},
\end{aligned}
\end{equation*}
with the uncertain scalar $\xi$ uniformly distributed over the interval $[-1, 1]$. 
Three $\mathcal{H}_\infty$ SOF control synthesis methods are implemented for comparison:
(\romannum{1}) worst-case robust control synthesis;
(\romannum{2}) the nominal PCE-based control synthesis in \eqref{eq:prob1_mi}; and (\romannum{3}) the proposed robust PCE-based control synthesis in Section \ref{sect:rb_pce_syn}.

The worst-case robust controller accounts for the polytopic uncertainty $\xi^3 \in [-1,1]$ by solving \cite{geromel2007H2}
\begin{equation*}\label{eqn:wc_syn}
\begin{aligned}
& \min_{\mathbf{P},\mathbf{K},\gamma} \;  \gamma \\
& \text{s.t.} \;
\begin{bmatrix}
\text{He}\{\mathbf{P} \mathbf{A}_{\text{cl}}(\xi_i)\}
& \mathbf{P} \mathbf{B}_{\text{cl}}(\xi_i) & \mathbf{C}_{\text{cl}}^\top(\xi_i) \\
\mathbf{B}_{\text{cl}}^\top(\xi_i) \mathbf{P} & 
-\gamma \mathbf{I} & \mathbf{D}_{\text{cl}}^\top(\xi_i) \\
\mathbf{C}_{\text{cl}}(\xi) & \mathbf{D}_{\text{cl}}(\xi_i) & -\gamma \mathbf{I}
\end{bmatrix} < 0, \\ 
&\qquad i=1,2,
\end{aligned}
\end{equation*}
with $\mathbf{A}_{\text{cl}}$, $\mathbf{B}_{\text{cl}}$, $\mathbf{C}_{\text{cl}}$, and $\mathbf{D}_{\text{cl}}$ defined in 
\eqref{eq:clsysmatdef}. 
The parameters $\{ \xi_i \}$ are set to be $\xi_1 = -1$ and $\xi_2=1$ such that 
$\{ \mathbf{A}_{\text{cl}}(\xi_i), \mathbf{B}_{\text{cl}}(\xi_i),
\mathbf{C}_{\text{cl}}(\xi_i), \mathbf{D}_{\text{cl}}(\xi_i) \}$ defines a polytope as an overbounding uncertainty set, i.e.,
\begin{align*}
\begin{bmatrix}
\mathbf{A}_{\text{cl}}(\xi) &
\mathbf{B}_{\text{cl}}(\xi) \\
\mathbf{C}_{\text{cl}}(\xi) & 
\mathbf{D}_{\text{cl}}(\xi)
\end{bmatrix} = 
\sum_{i=1}^{2} \beta_i 
\begin{bmatrix}
\mathbf{A}_{\text{cl}}(\xi_i) &
\mathbf{B}_{\text{cl}}(\xi_i) \\
\mathbf{C}_{\text{cl}}(\xi_i) & 
\mathbf{D}_{\text{cl}}(\xi_i)
\end{bmatrix}
\end{align*}
with $\sum_{i=1}^{2} \beta_i = 1$ and $\beta_i \geq 0$.  
The TOMLAB/PENBMI solver \cite{Holm2006} is used to solve the BMI optimization which produces the controller $\mathbf{K}_{\text{wc}} = 
\begin{bmatrix}
-0.1281 & -9.4664
\end{bmatrix}$.

\begin{figure}
	\centering
	\includegraphics[width=0.97\linewidth]{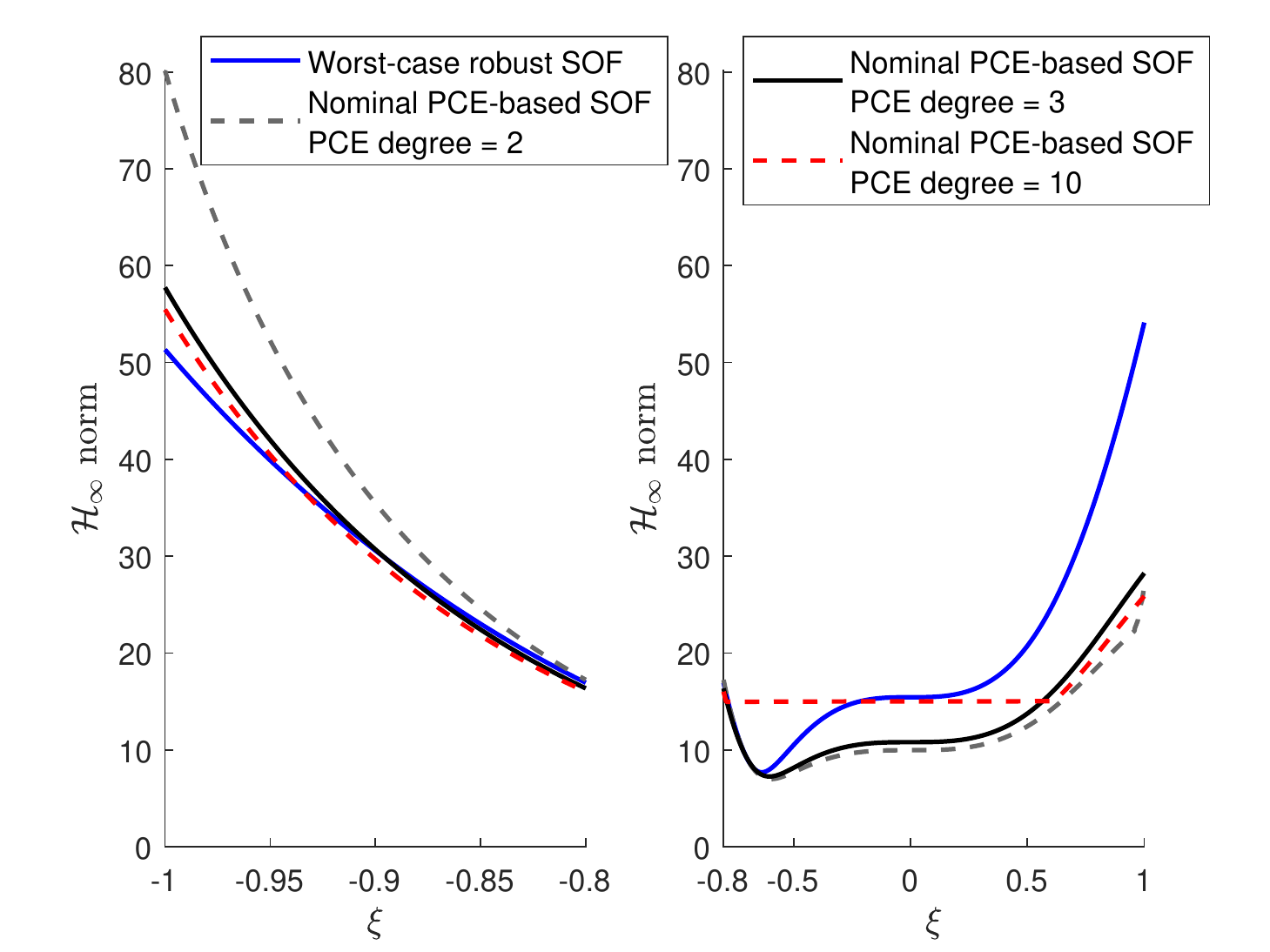}
	\caption{Distributions of $\mathcal{H}_\infty$ norms over $\xi \in [-1,1]$ generated by worst-case robust control and nominal PCE-based controls.}
	\label{fig:hinf_distr_1}
\end{figure}

To explicitly address the probabilistic uncertain parameter $\xi$, 
the nominal PCE-based control synthesis \eqref{eq:prob1_mi} is applied with different PCE degrees from 1 to 10.
Post-analysis shows that the obtained control gain does not robustly stabilize the closed-loop system if the adopted PCE degree is less than 2. 
Fig. \ref{fig:hinf_distr_1} depicts the distributions of $\mathcal{H}_\infty$ norms obtained from the worst-case robust control and the PCE-based control for 1000 values for $\xi$ uniformly distributed within $[-1, 1]$.
The nominal PCE-based control synthesis using PCEs with degrees 2, 3, and 10 gives $\mathbf{K}_\text{PCE}= 
\begin{bmatrix}
1.8539 & -27.4996
\end{bmatrix}$,
$\mathbf{K}_\text{PCE}= 
\begin{bmatrix}
1.5298 & -28.6719
\end{bmatrix}$, and 
$\mathbf{K}_\text{PCE} = 
\begin{bmatrix}
5.1988 & -74.7948
\end{bmatrix}$, respectively.
The nominal PCE-based controls achieve much better averaged performance than worst-case robust control, at the cost of larger worst-case $\mathcal{H}_\infty$ norms when $\xi$ approaches $-1$. 
As the adopted PCE degree increases to 10, the PCE truncation error becomes smaller, and the worst-case $\mathcal{H}_\infty$ norm decreases accordingly.

Without increasing the PCE degree, the robust PCE-based control synthesis \eqref{eq:prob2_mi} introduces a robustifying parameter $\rho^2$ to address the PCE truncation errors.
To illustrate this point, we consider low PCE degrees 1, 2, and 3. 
Firstly, the bisection algorithm described at the end of Section \ref{sect:rb_pce_syn} is applied to determine the lower bound $\rho_{\min}^2$ for $\rho^2$ that ensures closed-loop stability. The obtained lower bound $\rho_{\min}^2$ is 0.0027, 0, and 0 for the synthesis problem \eqref{eq:prob2_mi} with PCE degrees 1, 2, and 3, respectively. In the following, the robust PCE-based control synthesis \eqref{eq:prob2_mi} with the $2^\text{nd}$-degree PCE is solved with different values of $\rho^2$.
In this case, the synthesis problem \eqref{eq:prob2_mi} has 25 decision variables, which is significantly fewer than 256 decision variables in \eqref{eq:prob1_mi} using a $10^\text{th}$-degree PCE.
The resulting distributions of $\mathcal{H}_\infty$ norms are compared to the worst-case robust control and the nominal PCE-based control in Fig. \ref{fig:hinf_distr_1}.
Compared to the nominal PCE-based control with $2^\text{nd}$-degree PCEs, the robust PCE-based control with $2^\text{nd}$-degree PCEs and $\rho^2=0.0036$ achieves both a much smaller worst-case $\mathcal{H}_\infty$ norm and a smaller averaged $\mathcal{H}_\infty$ norm. 
By just increasing $\rho^2$ to 0.0225, the worst-case $\mathcal{H}_\infty$ norm is further reduced to be about 26\% smaller than for worst-case robust synthesis.
However, as $\rho^2$ increases from 0.0036 to 0.0225, a more conservative bound is used to quantify the effect of the PCE truncation errors, which leads to an increase in the averaged $\mathcal{H}_\infty$ norm from 14.6731 to 16.4820. For a detailed comparison, Table \ref{tab:norm} lists the worst-case and averaged $\mathcal{H}_\infty$ norms obtained by different methods and tuning parameters.

\begin{figure}
	\centering
	\includegraphics[width=0.97\linewidth]{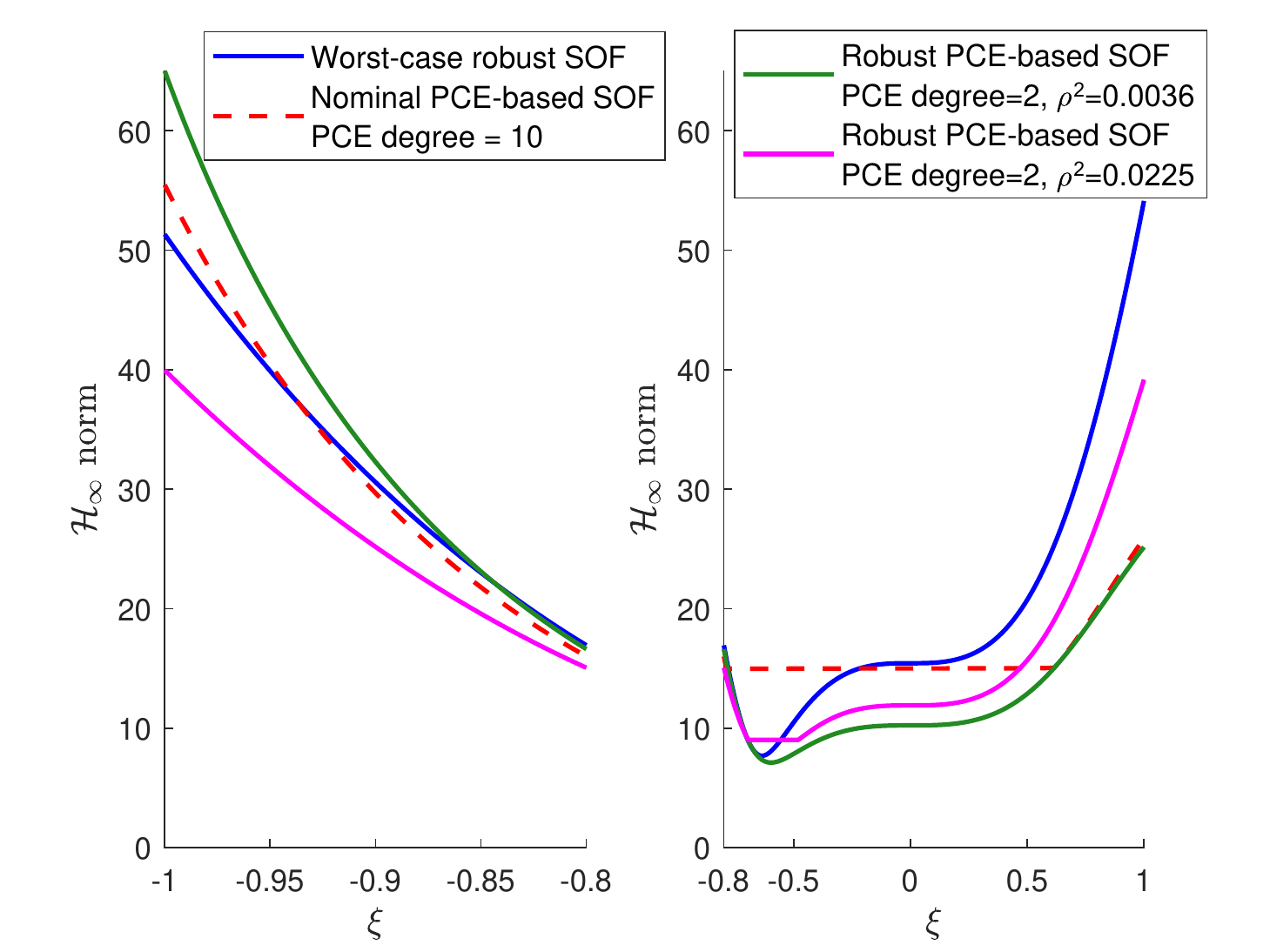}
	\caption{Distributions of $\mathcal{H}_\infty$ norms over $\xi \in [-1,1]$ generated by worst-case robust control, nominal PCE-based control, and robust PCE-based controls.}
	\label{fig:hinf_distr_2}
\end{figure}

\begin{table}
	\centering
	\caption{Worst-case and averaged $\mathcal{H}_\infty$ norms achieved by different synthesis methods ($p$ represents the PCE degree, and $\rho$ is the robustifying parameter in the robust PCE-based synthesis)}
	\begin{tabular}{ lSS } 
		\toprule
		{\multirow{2}{*}{Synthesis method}} & {Worst-case}  & {Averaged} \\
		& {$\mathcal{H}_\infty$ norm} & {$\mathcal{H}_\infty$ norm} \\
		\midrule 
		Worst-case robust SOF & 54.1316 & 21.0501 \\ \addlinespace[0.5em]
		Nominal PCE-based SOF & & \\ 
		$\quad p=2$ & 80.1360 & 14.7713 \\ 
		$\quad p=3$ & 57.7491 & 15.1790 \\ 
		$\quad p=10$ & 55.4751 & 17.7026 \\ \addlinespace[0.5em]
		\multicolumn{3}{l}{Robust PCE-based SOF} \\  
		$\quad p=2, \rho^2=0.0001$ & 79.3716 & 14.7584 \\ 
		$\quad p=2, \rho^2=0.0036$ & 65.0046 & 14.6731 \\ 
		$\quad p=2, \rho^2=0.0225$ & 39.9650 & 16.4820 \\ \addlinespace[0.5em]
		Nominal PCE-based SOF  & & \\
		using the PCE-transformed system \eqref{eq:clsys_comp} & & \\
		$\quad p=2$ & 168.3335 & 17.5934 \\ 
		$\quad p=3$ & 67.9149 & 17.2158 \\ 
		$\quad p=10$ & 55.4647 & 17.7024 \\ 
		\bottomrule
	\end{tabular}
	\label{tab:norm}
\end{table}

\begin{figure}
	\centering
	\includegraphics[width=0.9\linewidth]{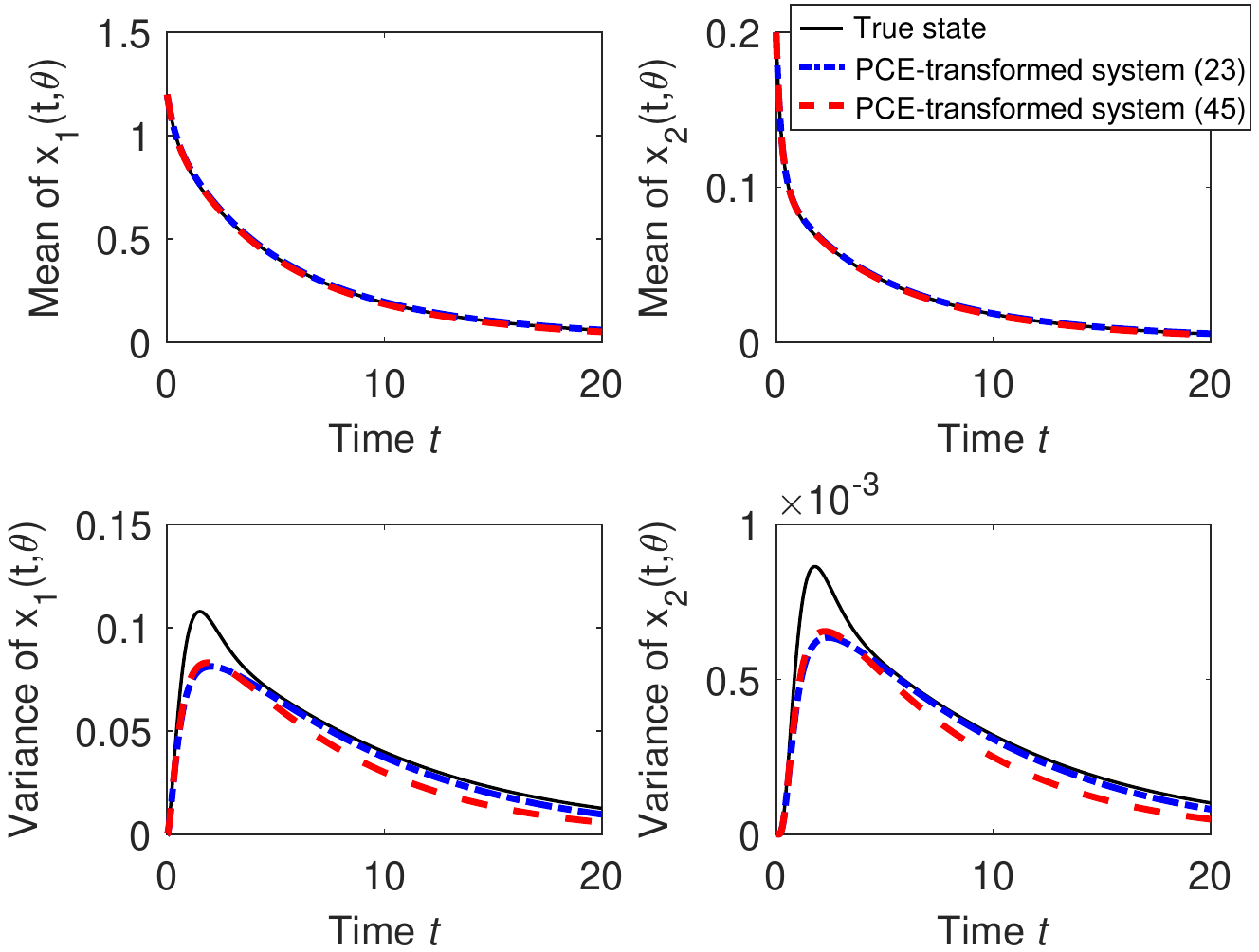}
	\caption{State means and variances obtained from PCE-transformed systems \eqref{eq:pce_dyn} and \eqref{eq:clsys_comp} compared to the true system state, with zero disturbances $\mathbf{w}$, $2^\text{nd}$-degree PCEs, and a SOF control gain $\mathbf{K}= [1.8539  \;\; -27.4996]$.}
	\label{fig:pce_err}
\end{figure}

The results of nominal PCE-based synthesis using the PCE-transformed system \eqref{eq:clsys_comp} as in \cite{Wan2018ACC} are also included in Table \ref{tab:norm}, in order to illustrate 
the benefit of the proposed closed-loop PCE transformation in Section \ref{sect:pce_cldyn}.
To compare the approximation errors, Fig. \ref{fig:pce_err} depicts the mean and variance of state trajectories produced by the true system and the two different PCE-transformed systems \eqref{eq:Xa} and \eqref{eq:clsys_comp}, with zero disturbance $\mathbf{w}$, $2^{\text{nd}}$-degree PCEs, and a SOF control gain $\mathbf{K}= \begin{bmatrix}
1.8539  & -27.4996
\end{bmatrix}$. 
The true state statistics are obtained from Monte Carlo simulations, while the state means and variances from the PCE-transformed systems are computed using \eqref{eq:mean_var}--\eqref{eq:pce_truncated}.
The proposed PCE-transformed closed-loop dynamics \eqref{eq:Xa} have higher accuracy than the PCE-transformed system \eqref{eq:clsys_comp}, although both have good approximation to the mean of the true system states.
This observation is consistent with the reasons explained in Appendix \ref{app:comp}, 
The proposed closed-loop PCE transformation results in significantly improved control performance when the PCE degree is 2 or 3, as shown in Table \ref{tab:norm}. 
As the PCE degree increases to 10, the difference between the nominal PCE-transformed systems proposed in this paper and that reported in \cite{Wan2018ACC} become negligible, which result in almost identical worst-case and averaged $\mathcal{H}_\infty$ norms.

\section{Conclusion}\label{sect:conclusion}

A robust PCE-based $\mathcal{H}_\infty$ SOF control synthesis method is presented to address probabilistic time-invariant parametric uncertainties. 
A closed-loop PCE-based transformation is proposed to achieve smaller transformation errors compared to reported PCE-based transformations in the literature.
The effect of PCE truncation errors is captured by a norm-bounded LDI to formulate a robust PCE-based synthesis method. 
The proposed approach allows the use of relatively low-degree PCEs by tuning a robustifying parameter.
In contrast, most existing PCE-based control design methods rely on using high-degree PCEs to alleviate the effect of truncation errors, which results in much higher computational cost.




\section*{APPENDIX}
\newcommand{\labelsubseccounter}[1]{
	\renewcommand\thesubsection{\Alph{subsection}}
	\addtocounter{subsection}{-1}
	\refstepcounter{subsection}
	\label{#1}
	\renewcommand\thesubsection{\thesection.\arabic{subsection}}
}

\subsection{Proof of Proposition \ref{prop:Acl_Bcl}}\labelsubseccounter{app:Acl_Bcl_pc}
According to \eqref{eq:clsysmatdef}, \eqref{eqn:innerproduct}, \eqref{eq:A_pc}, \eqref{eq:barB_pc}, \eqref{eq:hatC_pc}, and \eqref{eq:hatK_pc}, it follows that 
\begin{equation*}\label{eq:barAcl_pc_pf}
\begin{aligned}
\mathcal{\bar A}_\text{cl} &= \mathbb{E}_{\boldxi} \{ \Phi(\boldxi)\mathbf{A}_\text{cl}(\boldxi)\Phi^\top(\boldxi) \} \\
&= \mathcal{A} + \mathbb{E}_{\boldxi} \{ \Phi(\boldxi) \mathbf{B}(\boldxi)\mathbf{K}\mathbf{C}(\boldxi) \Phi^\top(\boldxi) \} \\
&= \mathcal{A} + \mathbb{E}_{\boldxi} \left\{ \left( \sum_{i=0}^{N_q} \mathbf{\hat B}_i \phi_i(\boldxi) \right) \mathbf{K} \left( \sum_{j=0}^{N_q} \mathbf{\hat C}_j \phi_j(\boldxi) \right) \right\} \\
&= \mathcal{A} + \mathbb{E}_{\boldxi} \left\{ \sum_{i=0}^{N_q} \sum_{j=0}^{N_q} \mathbf{\hat B}_i \mathbf{K} \mathbf{\hat C}_j \phi_i(\boldxi) \phi_j(\boldxi)\right\} \\
&= \mathcal{A} + \sum_{i=0}^{N_q} \mathbf{\hat B}_i \mathbf{K} \mathbf{\hat C}_i 
=\mathcal{A} + \mathcal{\bar B} \mathcal{\bar K} \mathcal{\bar C}.
\end{aligned}
\end{equation*}
The proof of $\mathcal{\bar B}_\text{cl}=\mathcal{B}_\mathbf{w} + \mathcal{\bar B}\mathcal{\bar K}\mathcal{\bar D}_\mathbf{w}$ follows similar procedures as above.

\subsection{Proof of Proposition \ref{prop:z}}\labelsubseccounter{app:z}
Since $\mathbb{E}_{\boldxi}\{ \left\| \mathbf{z}_\text{nom}(t,\boldxi) \right\|_{\mathcal{L}_2}^2 \}$ can be expressed as
\begin{equation*}
\begin{aligned}
\mathbb{E}_{\boldxi}\{ \left\| \mathbf{z}_\text{nom}(t,\boldxi) \right\|_{\mathcal{L}_2}^2 \}
&= \int_{0}^{\infty}
\begin{bmatrix}
\mathbf{X}(t) \\ \mathbf{w}(t)
\end{bmatrix}^\top 
\begin{bmatrix}
\Gamma_1 & \Gamma_2^\top \\
\Gamma_2 & \Gamma_3
\end{bmatrix}
\begin{bmatrix}
\mathbf{X}(t) \\ \mathbf{w}(t)
\end{bmatrix} \text{d}t,
\end{aligned}
\end{equation*}
the proof of Proposition \ref{prop:z} is equivalent to the proof of 
\begin{align}
\Gamma_1 &= \mathbb{E}_{\boldxi} \{ \Phi_\mathbf{x}(\boldxi) \mathbf{C}_\text{cl}^\top(\boldxi) \mathbf{C}_\text{cl}(\boldxi)  \Phi_\mathbf{x}^\top(\boldxi) \} =\mathcal{\bar C}_\text{cl}^\top \mathcal{\bar C}_\text{cl}, \label{eq:Gamma1}\\
\Gamma_2 &= \mathbb{E}_{\boldxi} \{ \mathbf{D}_\text{cl}^\top(\boldxi) \mathbf{C}_\text{cl}(\boldxi)  \Phi_\mathbf{x}^\top(\boldxi) \} = \mathcal{\bar D}_\text{cl}^\top \mathcal{\bar C}_\text{cl}, \label{eq:Gamma2}\\
\Gamma_3 &= \mathbb{E}_{\boldxi} \{ \mathbf{D}_\text{cl}^\top(\boldxi) \mathbf{D}_\text{cl}(\boldxi) \} = \mathcal{\bar D}_\text{cl}^\top \mathcal{\bar D}_\text{cl}.  \label{eq:Gamma3}
\end{align}

According to the definition of $\Phi_\mathbf{x}^\top(\boldxi)$ in \eqref{eq:vecs_pce}, the following equations hold: 
\begin{align}
\mathbf{C}_\mathbf{z} \Phi_\mathbf{x}^\top(\boldxi) &= \Phi_\mathbf{z}^\top(\boldxi) \mathcal{C}_\mathbf{Z}, \label{eq:Cz}\\
\mathbf{K}^\top \mathbf{D}_\mathbf{z}^\top \mathbf{C}_\mathbf{z} \Phi_\mathbf{x}^\top(\boldxi) &= \Phi_\mathbf{y}^\top(\boldxi) \mathcal{K}^\top \mathcal{D}_\mathbf{Z}^\top \mathcal{C}_\mathbf{Z}, \label{eq:Phiy}
\end{align}
with $\Phi_\mathbf{y}(\boldxi)$ and $\Phi_\mathbf{z}(\boldxi)$ defined similarly to $\Phi_\mathbf{x}(\boldxi)$,
\begin{equation}\label{eq:K_Dz_pc}
\mathcal{K} = \mathbf{I}_{N_p+1} \otimes \mathbf{K}, \;
\mathcal{D}_\mathbf{Z} = \mathbf{I}_{N_p+1} \otimes \mathbf{D}_\mathbf{z}. 
\end{equation}
From \eqref{eq:hatCi}, define $\mathcal{C}$ as
\begin{align}\label{eq:C_pc}
\mathcal{C} &= \mathbb{E}_{\boldxi}\{\Phi_\mathbf{y}(\boldxi)\mathbf{C}(\boldxi)\Phi_\mathbf{x}^\top(\boldxi)\} = \mathbb{E}_{\boldxi}\left\{\Phi_\mathbf{y}(\boldxi) \sum_{i=0}^{N_q} \mathbf{\hat C}_i \phi_i(\boldxi) \right\} \nonumber \\
&= \begin{bmatrix}
\mathbf{\hat C}_0^\top & \mathbf{\hat C}_1^\top & \cdots & \mathbf{\hat C}_{N_p}^\top 
\end{bmatrix}^\top\!.
\end{align}
Since the degree of $\mathbf{C}(\boldxi)\Phi_\mathbf{x}^\top(\boldxi)$ is higher than for $\Phi_\mathbf{x}^\top(\boldxi)$, i.e., $q>p$ and $N_q>N_p$, it follows from \eqref{eq:C_pc} and \eqref{eq:K_Dz_pc} that $\mathcal{\bar C}$ in \eqref{eq:hatC_pc}, $\mathcal{\bar K}$ in \eqref{eq:hatK_pc} and $\mathcal{\bar D}_\mathbf{Z}$ in \eqref{eq:hatDz_pc} can be partitioned as
\begin{equation}\label{eq:Cbar_partition}
\mathcal{\bar C} = \begin{bmatrix}
\mathcal{C} \\
\mathcal{C}_1
\end{bmatrix}, \;
\mathcal{\bar K} = \begin{bmatrix}
\mathcal{K} & \mathbf{0} \\
\mathbf{0} & \mathcal{K}_1
\end{bmatrix},\;
\mathcal{\bar D}_\mathbf{Z} = \begin{bmatrix}
\mathcal{D}_\mathbf{Z} & \mathbf{0} \\
\mathbf{0} & \mathcal{D}_{\mathbf{Z},1}
\end{bmatrix}.
\end{equation}
With the above equations, \eqref{eq:clsysmatdef}, \eqref{eq:barCZ_pc}--\eqref{eq:hatDz_pc}, \eqref{eq:Cz}, and \eqref{eq:Phiy}, it can be derived that
\begin{align*}
\Gamma_1 &= \mathbb{E}_{\boldxi} \{ \Phi_\mathbf{x}(\boldxi) \mathbf{C}_\text{cl}^\top(\boldxi) \mathbf{C}_\text{cl}(\boldxi)  \Phi_\mathbf{x}^\top(\boldxi) \} = \Gamma_{11} + \Gamma_{12} + \Gamma_{13}
\end{align*}
with 
\begin{align*}
\Gamma_{11} &= \mathbb{E}_{\boldxi} \{ \Phi_\mathbf{x}(\boldxi) \mathbf{C}_\mathbf{z}^\top \mathbf{C}_\mathbf{z} \Phi_\mathbf{x}^\top(\boldxi) \}
= \mathbb{E}_{\boldxi} \{ \mathcal{C}_\mathbf{Z}^\top \Phi_\mathbf{z}(\boldxi) \Phi_\mathbf{z}^\top(\boldxi) \mathcal{C}_\mathbf{Z} \} \\
&= \mathcal{C}_\mathbf{Z}^\top \mathbb{E}_{\boldxi} \{ \Phi_\mathbf{z}(\boldxi) \Phi_\mathbf{z}^\top(\boldxi) \} \mathcal{C}_\mathbf{Z} 
= \mathcal{C}_\mathbf{Z}^\top \mathcal{C}_\mathbf{Z} 
= \mathcal{\bar C}_\mathbf{Z}^\top \mathcal{\bar C}_\mathbf{Z}, \\
\Gamma_{12} &= \text{He}\left\{ \mathbb{E}_{\boldxi}\{ \Phi_\mathbf{x}(\boldxi)\mathbf{C}^\top (\boldxi) \mathbf{K}^\top \mathbf{D}_\mathbf{z}^\top \mathbf{C}_\mathbf{z} \Phi_\mathbf{x}^\top(\boldxi) \} \right\} \\
&= \text{He}\left\{ \mathbb{E}_{\boldxi}\{ \Phi_\mathbf{x}(\boldxi)\mathbf{C}^\top (\boldxi) \Phi_\mathbf{y}^\top(\boldxi) \} \mathcal{K}^\top \mathcal{D}_\mathbf{Z}^\top \mathcal{C}_\mathbf{Z} \right\} \\
&= \text{He}\{ \mathcal{C}^\top \mathcal{K}^\top \mathcal{D}_\mathbf{Z}^\top \mathcal{C}_\mathbf{Z} \}
= \text{He}\{ \mathcal{\bar C}^\top \mathcal{\bar K}^\top \mathcal{\bar D}_\mathbf{Z}^\top \mathcal{\bar C}_\mathbf{Z} \}, \\
\Gamma_{13} &= \mathbb{E}_{\boldxi} \{ \Phi_\mathbf{x}(\boldxi) \mathbf{C}^\top(\boldxi) \mathbf{K}^\top \mathbf{D}_\mathbf{z}^\top \mathbf{D}_\mathbf{z} \mathbf{K} \mathbf{C}(\boldxi) \Phi_\mathbf{x}^\top(\boldxi) \} \\
&= \mathbb{E}_{\boldxi} \left\{ \sum_{i=0}^{N_q}\sum_{j=0}^{N_q} \mathbf{\hat C}_i^\top  \mathbf{K}^\top \mathbf{D}_\mathbf{z}^\top \mathbf{D}_\mathbf{z} \mathbf{K} \mathbf{\hat C}_i \phi_i(\boldxi) \phi_j(\boldxi)  \right\} \\
&= \sum_{i=0}^{N_q} \mathbf{\hat C}_i^\top  \mathbf{K}^\top \mathbf{D}_\mathbf{z}^\top \mathbf{D}_\mathbf{z} \mathbf{K} \mathbf{\hat C}_i  
= \mathcal{\bar C}^\top \mathcal{\bar K}^\top \mathcal{\bar D}_\mathbf{Z}^\top \mathcal{\bar D}_\mathbf{Z} \mathcal{\bar K} \mathcal{\bar C}.
\end{align*}
The above expressions of $\Gamma_{11}$, $\Gamma_{12}$, and $\Gamma_{13}$ prove that $\Gamma_1 = \mathcal{\bar C}_\text{cl}^\top \mathcal{\bar C}_\text{cl}$ with $\mathcal{\bar C}_\text{cl}$ defined in \eqref{eq:Ccl_Dcl}.
The proofs of \eqref{eq:Gamma2} and \eqref{eq:Gamma3} follow 
similar procedures and so are omitted.

\subsection{Comparison with existing PCE-based approximations}
\labelsubseccounter{app:comp}
Firstly, the two PCE-transformed closed-loop systems for comparison are summarized. The first one is \eqref{eq:Xa}, while the second one is given in \cite{Fisher2009,Shen2017,Wan2018ACC}, and briefly reviewed below  without detailed derivations. 
By applying standard Galerkin projection, the four equations in \eqref{eqn:open_sys} and (\ref{eqn:sofcontroller}) are separately transformed into
\begin{subequations}\label{eq:pce_open}
\begin{align}
\mathbf{\dot X}_b (t) &= \mathcal{A}\mathbf{X}_b (t) + \mathcal{B}\mathbf{U}_b(t) + \mathcal{B}_\mathbf{w}\mathbf{w}(t), \label{eqn:transformed_dyn} \\
\mathbf{Z}_b (t) &= \mathcal{C}_\mathbf{Z}\mathbf{X}_b(t) + \mathcal{D}_{\mathbf{Z}\mathbf{w}} \mathbf{w}(t) + \mathcal{D}_\mathbf{Z}\mathbf{U}_b(t), \label{eqn:transformedoutput1} \\
\mathbf{Y}_b(t) &= \mathcal{C}\mathbf{X}_b(t) + \mathcal{D}_\mathbf{w} \mathbf{w}(t),  \label{eqn:transformedmeasurement}  \\
\mathbf{U}_b (t) &= \mathcal{K} \mathbf{Y}_b(t), \label{eqn:transformed_sofcontroller}
\end{align}
\end{subequations}
where 
the matrices $\mathcal{B}$ and $\mathcal{D}_\mathbf{w}$ are defined as 
\begin{align}\label{eq:B_Dw_pc}
\mathcal{B} = \mathbb{E}_{\boldxi}\{\Phi_\mathbf{x}(\boldxi)\mathbf{B}(\boldxi)\Phi_\mathbf{u}^\top(\boldxi)\} \text{ and } 
\mathcal{D}_\mathbf{w} = \mathbb{E}_{\boldxi}
\{ \Phi_\mathbf{y}(\boldxi) \mathbf{D}_\mathbf{w}(\boldxi) \}, 
\end{align}
with $\Phi_\mathbf{u}(\boldxi)$ and $\Phi_\mathbf{y}(\boldxi)$ defined similarly to $\Phi_\mathbf{x}(\boldxi)$ in \eqref{eq:vecs_pce}, 
$\mathcal{C}_\mathbf{Z}$,  
$\mathcal{D}_{\mathbf{Z}\mathbf{w}}$, 
$\mathcal{D}_\mathbf{Z}$, 
$\mathcal{K}$, and
$\mathcal{C}$ are defined in \eqref{eq:Cz_Dzw_pc}, \eqref{eq:K_Dz_pc}, \eqref{eq:C_pc}, respectively.
Similarly to \eqref{eq:Xa}, the error terms introduced in Galerkin projection are not included in \eqref{eq:pce_open}, hence $\mathbf{U}_b$,
$\mathbf{Y}_b$, and $\mathbf{Z}_b$ are approximates of the PCE coefficient vectors of the control input, measured system output, and controlled system output, respectively.
Combining \eqref{eqn:transformed_dyn}--\eqref{eqn:transformed_sofcontroller} leads to the PCE-transformed closed-loop system given in \cite{Shen2017,Wan2018ACC}:
\begin{equation}\label{eq:clsys_comp}
\begin{aligned}
\dot{\mathbf{X}}_b(t) &= \mathcal{A}_{\text{cl}}\mathbf{X}_b (t) + \mathcal{B}_{\text{cl}}\mathbf{w}(t), \\
\mathbf{Z}_b (t) &= \mathcal{C}_{\text{cl}} \mathbf{X}_b (t) + \mathcal{D}_{\text{cl}} \mathbf{w}(t),
\end{aligned}
\end{equation}
with
\begin{equation}\label{eq:clsysmat_comp}
\begin{aligned}
\mathcal{A}_\text{cl} &= \mathcal{A} + \mathcal{B}\mathcal{K}\mathcal{C}, \;
\mathcal{B}_\text{cl} = \mathcal{B}_\mathbf{w} + \mathcal{B}\mathcal{K}\mathcal{D}_\mathbf{w}, \\
\mathcal{C}_\text{cl} &= \mathcal{C}_\mathbf{Z} + \mathcal{D}_\mathbf{Z}\mathcal{K}\mathcal{C}, \;
\mathcal{D}_\text{cl} = \mathcal{D}_\mathbf{Zw} + \mathcal{D}_\mathbf{Z} \mathcal{K}\mathcal{D}_\mathbf{w}.
\end{aligned}
\end{equation}

Next, a relationship between the system matrices in \eqref{eq:Xa} and \eqref{eq:clsysmat_comp} is established.
Similarly to the partition of $\mathcal{\bar C}$ in \eqref{eq:Cbar_partition}, $\mathcal{\bar B}$ in \eqref{eq:barB_pc} and $\mathcal{\bar D}_\mathbf{w}$ in \eqref{eq:barDw_pc} are partitioned as 
\begin{equation}\label{eq:Bbar_partition}
\mathcal{\bar B} = \begin{bmatrix}
\mathcal{B} & \mathcal{B}_1
\end{bmatrix}\!, 
\mathcal{\bar D}_\mathbf{w} = \begin{bmatrix}
\mathcal{D}_\mathbf{w} \\
\mathcal{D}_{\mathbf{w},1}
\end{bmatrix}
\end{equation}
according to the definition of $\mathcal{B}$ and $\mathcal{D}_\mathbf{w}$ in \eqref{eq:B_Dw_pc}. 
By exploiting \eqref{eq:Cbar_partition} and \eqref{eq:Bbar_partition}, $\mathcal{\bar A}_\text{cl}$, $\mathcal{\bar B}_\text{cl}$ in \eqref{eq:Acl_Bcl_pc} and 
$\mathcal{\bar C}_\text{cl}$, $\mathcal{\bar D}_\text{cl}$ in \eqref{eq:Ccl_Dcl} can be rewritten as
\begin{equation}\label{eq:relation}
\begin{aligned}
\mathcal{\bar A}_\text{cl} &=
\mathcal{A}_\text{cl} + \mathcal{B}_1 \mathcal{K}_1 \mathcal{C}_1, \;
\mathcal{\bar B}_\text{cl} =
\mathcal{B}_\text{cl} + \mathcal{B}_1 \mathcal{K}_1 \mathcal{D}_{\mathbf{w},1}, \\
\mathcal{\bar C}_\text{cl} &=
\begin{bmatrix}
\mathcal{C}_\text{cl} \\
\mathcal{D}_{\mathbf{Z},1}
\mathcal{K}_1 \mathcal{C}_1
\end{bmatrix}\!,
\mathcal{\bar D}_\text{cl} =
\begin{bmatrix}
\mathcal{D}_\text{cl} \\
\mathcal{D}_{\mathbf{Z},1}
\mathcal{K}_1 \mathcal{D}_{\mathbf{w},1}
\end{bmatrix}.
\end{aligned}
\end{equation}
From \eqref{eq:barB_pc}--\eqref{eq:barDw_pc} and the matrix partitions in \eqref{eq:Cbar_partition} and \eqref{eq:Bbar_partition}, it can be seen that $\mathcal{B}_1$, $\mathcal{C}_1$, and $\mathcal{D}_{\mathbf{w},1}$ are determined by the high-degree PCE coefficients $\{ \mathbf{\hat B}_i, \mathbf{\hat C}_i, \mathbf{\hat D}_{\mathbf{w},i} \}_{i=N_p+1}^{N_q}$ in \eqref{eq:phi_BCDw}, 
but they are missing in the system matrices in \eqref{eq:clsys_comp}. Note that if the system matrix $\mathbf{B}$ in \eqref{eqn:open_sys} is independent of $\boldxi$, 
$\mathcal{B}_1$ defined in \eqref{eq:Bbar_partition} and \eqref{eq:Cbar_partition} becomes a zero matrix, 
which results in $\mathcal{\bar A}_\text{cl} = \mathcal{A}_\text{cl}$ and $\mathcal{\bar B}_\text{cl} = \mathcal{B}_\text{cl}$.
Similar consequences occur if $\mathbf{C}$ or $\mathbf{D}_\mathbf{w}$ is independent of $\boldxi$.

Since $\mathbf{X}_a(t)$ and $\mathbf{X}_b(t)$ in the PCE-transformed systems \eqref{eq:Xa} and \eqref{eq:clsys_comp} are approximates of PCE coefficients $\mathbf{X}(t)$ of the truce system state $\mathbf{x}(t,\boldxi)$, the state reconstruction error 
\begin{equation}\label{eq:app_err}
\mathbf{\tilde x}_\star(t,\boldxi) = \mathbf{x}(t,\boldxi) - \Phi_\mathbf{x}^\top (\boldxi) \mathbf{X}_\star (t)
\end{equation}
is used to compare the approximation errors of these PCE-transformed systems,
with $\star$ being $a$ or $b$. 
Such a state approximation error can be decomposed as 
\begin{equation}\label{eq:err_split}
\mathbf{\tilde x}_\star(t,\boldxi) = \mathbf{\tilde x}(t,\boldxi) + \Phi_\mathbf{x}^\top (\boldxi) \left( \mathbf{X}(t) - \mathbf{X}_\star(t) \right).
\end{equation}
with $\mathbf{\tilde x}(t,\boldxi)$ being the PCE truncation error defined in \eqref{eq:vecs_pce}.
The first term $\mathbf{\tilde x}(t,\boldxi)$ in \eqref{eq:err_split} is the orthogonal projection error determined by the true closed-loop system state and its PCE, thus is independent of the PCE-transformed system \eqref{eq:Xa} or \eqref{eq:clsys_comp}. 
In contrast, the second term $\Phi_\mathbf{x}^\top (\boldxi) \left( \mathbf{X}(t) - \mathbf{X}_\star(t) \right)$ in \eqref{eq:err_split} depends on the approximation adopted in deriving the PCE-transformed systems.
According to \eqref{eq:err_split}, the bound of $\mathbf{\tilde x}_\star(t,\boldxi)$ can be described as 
\begin{equation}\label{eq:bound_star}
\begin{aligned}
\mathbb{E}_{\boldxi} \{ \left\| \mathbf{\tilde x}_\star(t,\boldxi) \right\|_2^2 \} &\leq \mathbb{E}_{\boldxi} \{ \left\| \mathbf{\tilde x}(t,\boldxi)  \right\|_2^2 \} \\
&\quad + \mathbb{E}_{\boldxi} \{ \left\| \Phi_\mathbf{x}^\top (\boldxi) \left( \mathbf{X}(t) - \mathbf{X}_\star(t) \right)  \right\|_2^2 \} \\
&= \mathbb{E}_{\boldxi} \{ \left\| \mathbf{\tilde x}(t,\boldxi)  \right\|_2^2 \} + \left\| \mathbf{X}(t) - \mathbf{X}_\star(t) \right\|_2^2.
\end{aligned}
\end{equation}
To further derive an upper bound for \eqref{eq:bound_star}, the dynamics of 
$\mathbf{\tilde X}_\star(t) = \mathbf{X}(t) - \mathbf{X}_\star(t)$ is derived as
\begin{equation*}
	\begin{aligned}
		\mathbf{ \dot{\tilde X} }_a(t) &= \mathcal{\bar A}_\text{cl} \mathbf{\tilde X}_a(t) + \mathbf{\bar R_x}(t), \\
		\mathbf{ \dot{\tilde X} }_b(t) &= \mathcal{\bar A}_\text{cl} \mathbf{\tilde X}_b(t) + \mathcal{B}_1 \mathcal{K}_1 \mathcal{C}_1 \mathbf{X}_b(t) 
		+ \mathcal{B}_1 \mathcal{K}_1 \mathcal{D}_{\mathbf{w},1} \mathbf{w}(t) \\
		&\quad + \mathbf{\bar R_x}(t)
	\end{aligned}
\end{equation*}
according to \eqref{eq:pce_dyn}, \eqref{eq:Xa}, \eqref{eq:clsys_comp}, and \eqref{eq:clsysmat_comp}.
For the sake of brevity, assume $\mathbf{X}(0) = \mathbf{X}_a(0) = \mathbf{X}_b(0)$, then the upper bounds for the approximation errors $\mathbb{E}_{\boldxi} \{ \left\| \mathbf{\tilde x}_\star(t,\boldxi) \right\|_2^2 \}$ in \eqref{eq:bound_star} are expressed as
\begin{subequations}\label{eq:up_bound}
\begin{align}
	&\mathbb{E}_{\boldxi} \{ \left\| \mathbf{\tilde x}_a(t,\boldxi) \right\|_2^2 \}
	\leq \pi(t) + \left\| \int_0^t e^{ \mathcal{\bar A}_\text{cl} (t-\tau) } \mathbf{\bar R_x}(\tau)  d\tau \right\|_2^2, \label{eq:err_xa}\\
	&\mathbb{E}_{\boldxi} \{ \left\| \mathbf{\tilde x}_b(t,\boldxi) \right\|_2^2 \}
	\leq \pi(t) + \left\| \int_0^t e^{ \mathcal{\bar A}_\text{cl} (t-\tau) } \mathbf{\bar R_x}(\tau)  d\tau \right\|_2^2 \label{eq:err_xb}\\
	&\; +
	\left\| \int_0^t e^{ \mathcal{\bar A}_\text{cl} (t-\tau) } \mathcal{B}_1 \mathcal{K}_1 \left(  \mathcal{C}_1 \mathbf{X}_b(\tau) 
	+ \mathcal{D}_{\mathbf{w},1} \mathbf{w}(\tau)  \right) d\tau \right\|_2^2 \nonumber
\end{align}
\end{subequations}
for the PCE-transformed systems \eqref{eq:Xa} and \eqref{eq:clsys_comp}, respectively. 
Note that $\pi(t)$ in \eqref{eq:up_bound} represents the upper bound for the PCE truncation error 
$\mathbb{E}_{\boldxi} \{ \left\| \mathbf{\tilde x}(t,\boldxi) \right\|_2^2 \}$. The detailed expression of $\pi(t)$ has been given in literature such as \cite{audouze2016priori,muhlpfordt2018comments}, hence is not detailed here.

From \eqref{eq:bound_star} and \eqref{eq:up_bound}, it can be concluded that $\mathbb{E}_{\boldxi} \{ \left\| \mathbf{\tilde x}_a(t,\boldxi) \right\|_2^2 \}$ from our proposed PCE-transformed system \eqref{eq:Xa} has a smaller worst-case upper bound than $\mathbb{E}_{\boldxi} \{ \left\| \mathbf{\tilde x}_b(t,\boldxi) \right\|_2^2 \}$ from the PCE-transformed system in \cite{Fisher2009,Shen2017,Wan2018ACC}.

\subsection{Proof of Proposition \ref{prop:MN}}\labelsubseccounter{app:MN}
With Kronecker product, the first equation in \eqref{eq:xtilde2xpce} can be equivalently expressed as
\begin{equation}\label{eq:Meq}
\mathbf{\tilde x}(t,\boldxi) = \left( \left( \mathbf{X}^\top (t) \Phi_\mathbf{x}(\boldxi) \right) \otimes \mathbf{I}_{n_x} \right) 
\mathrm{vec} \left( \mathbf{M}(t,\boldxi) \right).
\end{equation}
For any given value of $t$ and $\boldxi$, the above equation \eqref{eq:Meq} must have a non-unique solution for $\mathrm{vec} \left( \mathbf{M}(t,\boldxi) \right)$
since the matrix $\left( \mathbf{X}^\top (t) \Phi_\mathbf{x}(\boldxi) \right) \otimes \mathbf{I}_{n_x}$ has full row-rank. The second equality of \eqref{eq:xtilde2xpce} exploits 
$\mathbf{M}(t,\boldxi) \Phi_\mathbf{x}^\top (\boldxi) = \Phi_\mathbf{x}^\top (\boldxi) \mathbf{N}(t,\boldxi)$  
according to the definition of $\Phi_\mathbf{x}^\top (\boldxi)$ in \eqref{eq:vecs_pce}.

\subsection{Proof of Theorem \ref{thm:rb_pce}}\labelsubseccounter{app:rb_pce}
By applying the Schur complement lemma, \eqref{eq:lmi_syn} is equivalent to 
\begin{equation}\label{eq:lmi_syn2}
\begin{aligned}
& 
\begin{bmatrix}
\mathrm{He}\{ \mathbf{P} \mathcal{\bar A}_\text{cl}  \}
& \mathbf{P}\mathcal{\bar B}_{\text{cl}}
& \mathbf{P} \mathcal{\bar A}_\text{cl} \\
\mathcal{\bar B}_{\text{cl}}^\top \mathbf{P} &
- \gamma_\text{rob} \mathbf{I} & \mathbf{0} \\
\mathcal{\bar A}_\text{cl}^\top \mathbf{P} & \mathbf{0} 
& \mathbf{0} 
\end{bmatrix}
+ \text{diag}(\tau \rho^2 \mathbf{I}, \mathbf{0}, -\tau \mathbf{I}) \\
& \qquad\quad + \gamma_\text{rob}^{-1}
\begin{bmatrix}
\mathcal{\bar C}_{\text{cl}}  &
\mathcal{\bar D}_{\text{cl}} & 
\mathcal{\bar C}_{\text{cl}}
\end{bmatrix}^\top 
\begin{bmatrix}
\mathcal{\bar C}_{\text{cl}}  &
\mathcal{\bar D}_{\text{cl}} & 
\mathcal{\bar C}_{\text{cl}}
\end{bmatrix} < 0.
\end{aligned}
\end{equation}

Define $\boldsymbol{\varphi}_\mathbf{x}(t)= \Delta_\mathbf{x}(t) \mathbf{X}_1(t)$ and $\chi(t) = \begin{bmatrix}
\mathbf{X}_1^\top(t) & \mathbf{w}^\top(t) & \boldsymbol{\varphi}_\mathbf{x}^\top(t)
\end{bmatrix}^\top$.
Let $V(t) = \mathbf{X}_1^\top(t) \mathbf{P} \mathbf{X}_1(t)$ represent the Lyapunov function. 
Left-multiplying \eqref{eq:lmi_syn2} by $\chi^\top(t)$ and right-multiplying \eqref{eq:lmi_syn2} by $\chi(t)$
results in
\begin{align*}
& \frac{\text{d}V}{\text{d} t}
+ \gamma_\text{rob}^{-1} \mathbf{Z}_\text{rob}^\top(t) \mathbf{Z}_\text{rob}(t)
- \gamma_\text{rob} \mathbf{w}^\top(t) \mathbf{w}(t) \\
&\qquad\qquad + \tau ( \rho^2 \mathbf{X}_1^\top(t) \mathbf{X}_1(t) - \boldsymbol{\varphi}_\mathbf{x}^\top(t) \boldsymbol{\varphi}_\mathbf{x}(t) )
< 0.
\end{align*}
Note that $\tau > 0$, and $\rho^2 \mathbf{X}_1^\top(t) \mathbf{X}_1(t) - \boldsymbol{\varphi}_\mathbf{x}^\top(t) \boldsymbol{\varphi}_\mathbf{x}(t) \geq 0$ for ${\Delta}_\mathbf{x}(t) \in \mathcal{F}_\mathbf{x}$.
According to the $\mathcal{S}$-procedure, the above inequality is equivalent to 
\begin{equation}\label{eq:dvdt}
\frac{\text{d}V}{\text{d} t}
+ \gamma_\text{rob}^{-1} \mathbf{Z}_\text{rob}^\top(t) \mathbf{Z}_\text{rob}(t) - \gamma_\text{rob} \mathbf{w}^\top(t) \mathbf{w}(t) < 0.
\end{equation}
With $\boldsymbol{\varphi}_\mathbf{x}(t)= \Delta_\mathbf{x}(t) \mathbf{X}_1(t)$, \eqref{eq:dvdt} can be rewritten as
$\chi^\top(t) \mathcal{P} \chi(t) < 0$ with 
\begin{align*}
\mathcal{P}=
\begin{bmatrix}
\text{He}\{\mathbf{P} \mathcal{\bar A}_\text{cl} ( \mathbf{I} +\Delta_\mathbf{x}(t))\}
& \mathbf{P} \mathcal{\bar B}_{\text{cl}} & 
( \mathbf{I} +\Delta_\mathbf{x}(t))^\top
\mathcal{\bar C}_{\text{cl}}^\top \\
\mathcal{\bar B}_{\text{cl}}^\top \mathbf{P} & 
-\gamma_\text{rob} \mathbf{I} & \mathcal{\bar D}_{\text{cl}}^\top \\
\mathcal{\bar C}_{\text{cl}} ( \mathbf{I} +\Delta_\mathbf{x}(t)) & \mathcal{\bar D}_{\text{cl}} & -\gamma_\text{rob} \mathbf{I}
\end{bmatrix}.
\end{align*}
Since $\chi^\top(t) \mathcal{P} \chi(t) < 0$ holds for arbitrary $\chi(t)$ and ${\Delta}_\mathbf{x}(t) \in \mathcal{F}_\mathbf{x}$, the quadratic stability and the $\mathcal{H}_\infty$ norm of the system \eqref{eq:pce_sys_unc} is proved according to the bounded real Lemma \cite{dullerud2013course}.

\subsection{Proof of Corollary \ref{thm:stable}}\labelsubseccounter{app:cor_stability}
The additive noise $\mathbf{w}(t)$ is set to zero for the internal stability analysis. The SOF controller obtained from solving \eqref{eq:prob2_mi} robustly stabilizes the PCE-transformed LDI \eqref{eq:pce_sys_unc}, hence 
$\mathbf{X}_1 (t)$ converges to zero with time. 
Since the trajectory of $\mathbf{X}(t)$ in the system \eqref{eq:pce_insert_dynall} belongs the trajectory set of $\mathbf{X}_1(t)$ in the LDI \eqref{eq:Fx}--\eqref{eq:pce_sys_LDI}, then we have $\mathbf{X}(\infty)=0$ implied by $\mathbf{X}_1(\infty) = 0$. As $\mathbf{N}(t,\boldxi)$ in \eqref{eq:xtilde2xpce} is bounded with probability 1 for all $t$, the truncation error $\mathbf{\tilde x}(t,\boldxi)$ approaches zero at almost every $\boldxi$ as $t$ goes to infinity. Therefore, we have 
\begin{equation}
	\begin{aligned}
		&\lim\limits_{t\rightarrow \infty} \mathbb{E}_{\boldxi} \{ \left\| \mathbf{x}(t,\boldxi) \right\|_2^2 \}
		=\lim\limits_{t\rightarrow \infty} \mathbb{E}_{\boldxi} \{ \left\| \mathbf{\tilde x}(t,\boldxi) + \Phi_\mathbf{x}^\top(\boldxi) \mathbf{X}(t) \right\|_2^2 \}  \\
		&\quad\quad \leq \lim\limits_{t\rightarrow \infty} \mathbb{E}_{\boldxi} \{ \left\| \mathbf{\tilde x}(t,\boldxi) \right\|_2^2 \} + \lim\limits_{t\rightarrow \infty} \left\| \mathbf{X}(t) \right\|_2^2 = 0
	\end{aligned}
\end{equation}
which proves the mean-square stability of the original closed-loop system \eqref{eq:clsys}.


%


\bibliographystyle{plain}
\bibliography{D:/GoogleDrive/MyRefs}

\begin{thebibliography}{10}

\bibitem{audouze2016priori}
C.~Audouze and P.~B. Nair.
\newblock A priori error analysis of stochastic {G}alerkin projection schemes
  for randomly parameterized ordinary differential equations.
\newblock {\em International Journal for Uncertainty Quantification},
  6(4):287--312, 2016.

\bibitem{boyarski2005robust}
S.~Boyarski and U.~Shaked.
\newblock Robust {$\mathcal{H}_\infty$} control design for best mean
  performance over an uncertain-parameters box.
\newblock {\em Systems \& Control Letters}, 54(6):585--595, 2005.

\bibitem{chesi2013exact}
G.~Chesi.
\newblock Exact robust stability analysis of uncertain systems with a scalar
  parameter via {LMIs}.
\newblock {\em Automatica}, 49:1083--1086, 2013.

\bibitem{chesi2009homo}
G.~Chesi, A.~Garulli, A.~Tesi, and A.~Vicino.
\newblock {\em Homogeneous Polynomial Forms for Robustness Analysis of
  Uncertain Systems}.
\newblock Springer-Verlag, 2009.

\bibitem{dullerud2013course}
G.~E. Dullerud and F.~Paganini.
\newblock {\em A Course in Robust Control Theory: a Convex Approach}.
\newblock Springer Science \& Business Media, 2013.

\bibitem{Fisher2009}
J.~Fisher and R.~Bhattacharya.
\newblock Linear quadratic regulation of systems with stochastic parameter
  uncertainties.
\newblock {\em Automatica}, 45:2831--2841, 2009.

\bibitem{geromel2007H2}
J.~C. Geromel, R.~H. Korogui, and J.~Bernussou.
\newblock $\mathcal{H}_2$ and $\mathcal{H}_\infty$ robust output feedback
  control for continuous time polytopic systems.
\newblock {\em IET Control Theory \& Applications}, 1(5):1541--1549, 2007.

\bibitem{hinrichsen1998stochastic}
D.~Hinrichsen and A.~J. Pritchard.
\newblock Stochastic {$H^{\infty}$}.
\newblock {\em SIAM Journal on Control and Optimization}, 36(5):1504--1538,
  1998.

\bibitem{Holm2006}
K.~Holmstr\"{o}m, A.~O. G\"{o}ran, and M.~M. Edvall.
\newblock {\em User's Guide for TOMLAB/PENOPT}.
\newblock Tomlab Optimization Inc.

\bibitem{Hsu2020design}
S.~Hsu and R.~Bhattacharya.
\newblock Design of linear parameter varying quadratic regulator in polynomial
  chaos framework.
\newblock {\em International Journal of Robust and Nonlinear Control},
  30(16):6661--6682, 2020.

\bibitem{lavaei2008robust}
J.~Lavaei and A.~G. Aghdam.
\newblock Robust stability of {LTI} systems over semialgebraic sets using
  sum-of-squares matrix polynomials.
\newblock {\em IEEE Transactions on Automatic Control}, 53(1):417--423, 2008.

\bibitem{Lucia2017}
S.~Lucia, J.~A. Paulson, R.~Findeison, and R.~D. Braatz.
\newblock On stability of stochastic linear systems via polynomial chaos
  expansions.
\newblock In {\em Proceedings of the 2017 American Control Conference}, pages
  5089--5094, Seattle, WA, 2017.

\bibitem{LeMa2010}
O.~P.~Le Ma\^{i}tre and O.~M. Knio.
\newblock {\em Spectral Methods for Uncertainty Quantification With
  Applications to Computational Fluid Dynamics}.
\newblock Springer, 2010.

\bibitem{muhlpfordt2018comments}
T.~M{\"u}hlpfordt, R.~Findeisen, V.~Hagenmeyer, and T.~Faulwasser.
\newblock Comments on truncation errors for polynomial chaos expansions.
\newblock {\em IEEE Control Systems Letters}, 2(1):169--174, 2018.

\bibitem{nandi2017poly}
S.~Nandi, V.~Migeon, T.~Singh, and P.~Singla.
\newblock Polynomial chaos based controller design for uncertain linear systems
  with state and control constraints.
\newblock {\em Journal of Dynamic Systems, Measurement, and Control},
  140(7):071009, 2017.

\bibitem{Paulson2015}
J.~A. Paulson, E.~Harinath, L.~C. Foguth, and R.~D. Braatz.
\newblock Nonlinear model predictive control of systems with probabilistic
  time-invariant uncertainties.
\newblock In {\em Proceedings of 5th IFAC Conference on Nonlinear Model
  Predictive Control}, pages 16--25, Seville, 2015.

\bibitem{paulson2018efficient}
J.~A. Paulson and A.~Mesbah.
\newblock An efficient method for stochastic optimal control with joint chance
  constraints for nonlinear systems.
\newblock {\em International Journal of Robust and Nonlinear Control}, 2018.
\newblock DOI:10.1002/rnc.3999.

\bibitem{petersen2014robust}
I.~R. Petersen and R.~Tempo.
\newblock Robust control of uncertain systems: classical results and recent
  developments.
\newblock {\em Automatica}, 50(5):1315--1335, 2014.

\bibitem{Shen2017}
D.~Shen, S.~Lucia, Y.~Wan, R.~Findeisen, and R.~D. Braatz.
\newblock Polynomial chaos-based {$\mathcal{H}_2$}-optimal static output
  feedback control of systems with probabilistic parameter uncertainties.
\newblock In {\em Proceedings of 20th IFAC World Congress}, pages 3595--3600,
  Toulouse, France, 2017.

\bibitem{Suli2003}
E.~S\"{u}li and D.~F. Mayers.
\newblock {\em An Introduction to Numerical Analysis}.
\newblock Cambridge University Press, 2003.

\bibitem{Tempo2013}
R.~Tempo, G.~Calafiore, and F.~Dabbene.
\newblock {\em Randomized Algorithms for Analysis and Control of Uncertain
  Systems with Applications}.
\newblock Springer-Verlag, London, 2013.

\bibitem{Wan2018mixed}
Y.~Wan and R.~D. Braatz.
\newblock Mixed polynomial chaos and worst-case synthesis approach to robust
  observer based linear quadratic regulation.
\newblock In {\em Proceedings of the 2018 American Control Conference}, pages
  6798--6803, Milwaukee, USA, 2018.

\bibitem{Wan2018ACC}
Y.~Wan, D.~E. Shen, S.~Lucia, R.~Findeisen, and R.~D. Braatz.
\newblock Robust static $\mathcal{H}_\infty$ output-feedback control using
  polynomial chaos.
\newblock In {\em Proceedings of the 2018 American Control Conference}, pages
  6804--6809, Milwaukee, USA, 2018.

\bibitem{Xiu2002}
D.~Xiu and G.~E. Karniadakis.
\newblock The {Wiener-Askey} polynomial chaos for stochastic differential
  equations.
\newblock {\em SIAM Journal of Scientific Computation}, 24:619--644, 2002.

\bibitem{yaesh2003probability}
I.~Yaesh, S.~Boyarski, and U.~Shaked.
\newblock Probability-guaranteed robust $\mathcal{H}_\infty$ performance
  analysis and state-feedback design.
\newblock {\em Systems \& Control Letters}, 48(5):351--364, 2003.

\end{thebibliography}

\end{document}